\documentclass{emulateapj-rtx4}
\usepackage{color}
\usepackage{wrapfig}
\usepackage{lscape}
\usepackage{amsmath}
\usepackage{pdflscape}
\usepackage{longtable}
\usepackage[version=3]{mhchem}
\usepackage[normalem]{ulem}

\newcommand{\feh}{$\mbox{[Fe/H]}$}

\begin{document}
\shorttitle{Carbon-Enhanced Metal-Poor Stars in Sculptor} 
\title{Detection of a Population of Carbon-Enhanced Metal-Poor Stars
  in the Sculptor Dwarf Spheroidal Galaxy\altaffilmark{*}}

\author{Anirudh Chiti\altaffilmark{1,2,*}, Joshua
  D. Simon\altaffilmark{3}, Anna Frebel\altaffilmark{1,2}, Ian
  B. Thompson\altaffilmark{3}, Stephen A. Shectman\altaffilmark{3},
  Mario Mateo\altaffilmark{4}, John I. Bailey, III\altaffilmark{5}, 
  Jeffrey D. Crane\altaffilmark{3},
  Matthew Walker\altaffilmark{6}}

\altaffiltext{*}{This paper includes data gathered with the 6.5 meter Magellan Telescopes located at Las Campanas Observatory, Chile.}
\altaffiltext{1}{Department of Physics and Kavli Institute for Astrophysics and 
                         Space Research,
                         Massachusetts Institute of Technology, 
                         Cambridge, MA 02139, USA; Email: \texttt{achiti@mit.edu}}
\altaffiltext{2}{JINA Center for the Evolution of the Elements, USA}                         
\altaffiltext{3}{Observatories of the Carnegie Institution 
			for Science, 813 Santa Barbara St., 
			Pasadena, CA 91101, USA}
\altaffiltext{4}{Department of Astronomy, University of Michigan, 
			Ann Arbor, MI 48109, USA}
\altaffiltext{5}{Leiden Observatory, Leiden University, P.O. Box 9513, 2300RA 						 Leiden, The Netherlands}
\altaffiltext{6}{McWilliams Center for Cosmology, 
			Department of Physics, Carnegie Mellon University,
			5000 Forbes Avenue, Pittsburg, PA 15213, USA}

\begin{abstract}
The study of the chemical abundances of metal-poor stars in dwarf galaxies provides a venue to constrain paradigms of chemical enrichment and galaxy formation. Here we present metallicity and carbon abundance measurements of 100 stars in Sculptor from medium-resolution ($R \sim$ 2000) spectra taken with the Magellan/Michigan Fiber System mounted on the Magellan-Clay 6.5m telescope at Las Campanas Observatory. We identify 24 extremely metal-poor star candidates ([Fe/H] $<$ $-$3.0) and 21 carbon-enhanced metal-poor (CEMP) star candidates. Eight carbon-enhanced stars are classified with at least 2$\sigma$ confidence and five are confirmed as such with follow-up $R\sim6000$ observations using the Magellan Echellette Spectrograph on the Magellan-Baade 6.5m telescope. We measure a CEMP fraction of 36\% for stars below [Fe/H] = $-$3.0, indicating that the prevalence of carbon-enhanced stars in Sculptor is similar to that of the halo ($\sim43\%$) after excluding likely CEMP-s and CEMP-r/s stars from our sample. However, we do not detect that any CEMP stars are strongly enhanced in carbon ([C/Fe] $>$ 1.0). The existence of a large number of CEMP stars both in the halo and in Sculptor suggests that some halo CEMP stars may have originated from accreted early analogs of dwarf galaxies.

\end{abstract}
\section{Introduction}
\label{sec:introduction}

The oldest stars in the Milky Way contain trace amounts of elements heavier than helium (or ``metals")
and measurements of their relative chemical abundances provide key constraints on the early phases of
chemical evolution \citep[e.g.][]{m+97, kcm+11}, galaxy formation \citep[e.g.][]{kb+02},
and the star-formation history (SFH) and initial mass function (IMF)
of their birth environment \citep[e.g.][]{bl+04}.
Studying metal-poor (MP) stars ([Fe/H] $< -1.0$, where [Fe/H] = $\log_{10}(N_{Fe}/N_H)_{\star} 
- \log_{10}(N_{Fe}/N_H)_\sun$) and in particular, extremely metal-poor (EMP)
stars ([Fe/H] $< -3.0$) in the Milky Way's dwarf satellite galaxies
effectively probes the aforementioned topics due to
the simpler dynamical and chemical evolution histories of dwarf galaxy systems
(see \citealt{tht+09} for a complete review).
Furthermore, dwarf galaxies have innate cosmological significance as they are hypothesized 
to be the surviving analogs of the potential building blocks of larger systems in hierarchical galaxy 
formation scenarios. Studying the most metal-poor stars in these systems
is a promising avenue to explore this intriguing potential connection.

While the specific relationship between dwarf galaxies and their ancient analogs
is not entirely understood, detailed abundance studies of the most metal-poor stars in
ultra-faint dwarf galaxies and classical dwarf spheroidal (dSph) galaxies
have shown some remarkable similarities between the chemical composition of EMP stars in dSphs 
and EMP stars in the halo of the Milky Way \citep{ch+09, ch+10,
kgb+09,fks+10,fsg+10, sfm+10, tjh+10, ngw+10, nyg+10, 
llb+11, gnm+13, fsk+14, kr+14, rk+14, sjf+15, jnm+15, jfs+16}. These results hint, at some level, of universality
in early chemical evolution and suggest that some of the
most metal-poor stars in
the Milky Way halo could have formed in dwarf galaxies. Because of the rarity of EMP stars, further identification and study of
these objects in any dwarf galaxy provides key information to further investigate these initial findings.

Chemically characterizing members of the Sculptor dSph galaxy has provided
insights on its chemical evolution and formation using 
high-resolution spectroscopy of red giant stars \citep{svt+03, tvs+03, gsw+05}. \citet{tih+04} found 
evidence for two stellar components in Sculptor, as also seen in other dSphs.
More recently, \citet{kgb+09} and the DART team \citep{bht+08, sht+10, rs+13} used samples of $\sim400-600$ Sculptor stars to derive the metallicity distribution function (MDF). 
Later, \citet{kcm+11} used the MDFs of Sculptor and other dSphs to investigate chemical 
evolution models.
Additional modeling of Sculptor by \citet{dts+12} showed evidence for extended star formation, and further modeling by \citet{rs+13} suggested the importance of dilution and metal-removal in chemical evolution scenarios.
Moreover, observations of a few individual EMP stars in Sculptor provided
the first evidence that low-metallicity stars
in dSphs are present and have chemical signatures matching those of EMP halo stars \citep{fks+10,tjh+10}.
Further studies of the S abundances of stars in Scuptor have shown similarities with the halo at lower metallicities \citep{sat+15}, and studies of Zn abundances have suggested complex nucleosynthetic origins for the element \citep{sts+17}.
Recently, work by \citet{sjf+15} and \citet{jnm+15} has indicated that EMP stars in Sculptor 
may have been enriched by just a handful of supernovae from the first generation of stars. 

The population of stars with [Fe/H] $<$ --2.5 in the Milky Way halo has long been 
known to include a large fraction enhanced in carbon \citep{bps+92, rbs+99, anr+02, ryan+03, bc+05, cst+05,
abc+07, pfb+14, fn+15}. 
This discovery led to the classification of 
carbon-enhanced metal-poor (CEMP) stars (metal-poor stars with [C/Fe] $>$ 0.7), within
which exist subdivisions contingent on the enhancements of r-process and/or s-process elements. 
Of those, CEMP-s and CEMP-rcd /s stars are readily explained as the products of binary mass
transfer from an asymptotic giant branch (AGB) companion \citep{ltb+05,han+16b}. However, stars that show [C/Fe] 
enhancement reflecting the chemical composition of their formative gas cloud,
as is thought to be the case for CEMP-r and CEMP-no stars, are the most 
useful in constraining theories of early chemical evolution.
Proposed mechanisms behind this early carbon enhancement include ``mixing and fallback''
SNe and massive rotating stars with large [C/Fe] yields, as discussed in e.g.,  \citet{nyb+13}.

Interestingly, the current sample of stars in Sculptor with [Fe/H] $<
-2.5$ from \citet{sht+13}, \citet{sjf+15}, and \citet{jnm+15}
contains no CEMP stars, contrary to expectations set by the high fraction of CEMP halo 
stars and earlier results that low-metallicity
chemical evolution appears to be universal. Only one CEMP-no star has been previously detected
in Sculptor \citep{sts+15}, with [Fe/H] = --2.03 and [C/Fe] $\sim$ 0.51, and only
three CEMP-s stars are known in the galaxy out of spectroscopic samples of hundreds of stars \citep{lbp+16,sdy+16}.
Under the assumptions that the ancient analogs of today's dwarf galaxies formed the Milky Way halo,
one would expect that dwarf galaxies should show carbon enhancement in their 
oldest stellar population as well. Earlier work detected a number of carbon-strong
stars in dSph galaxies, including Sculptor, but did 
not report individual metallicities for stars, precluding the characterization of these detected carbon-strong stars as
CEMP stars \citep{cnn+81, mcf+82, fbc+82, rw+83, aho+83, mc+83, alw+85, alw+86}. 
More recent searches in dSph galaxies \citep{llb+11, sss+13, sht+13,
sts+15, kgz+15, sks+17} have, however, detected only a handful of any category of CEMP stars. 

To investigate this apparent dearth of true CEMP stars, or CEMP-no stars, we surveyed Sculptor with the goal of identifying 
EMP star candidates and robustly characterizing its metal-poor population (Hansen et al., in prep). 
We conducted follow-up observations of the
most promising of these candidates to establish the low-metallicity tail of the MDF of 
Sculptor, and constrain the CEMP fraction in the system. 
In this paper, we present [Fe/H] and [C/Fe] measurements of the stars in our sample.
In Section~\ref{sec:obs}, we provide an overview of the target
selection and observations. In Sections~\ref{sec:Metallicity} and~\ref{sec:carbon}, we outline our methods
of obtaining [Fe/H] and [C/Fe] abundances for our sample. In Section~\ref{sec:additional},
we discuss additional measurements and considerations that are useful in analyzing our sample. We present
our results, discuss implications, and conclude in Section~\ref{sec:discussion}.\\\\

\section{Observations and Data Reduction}
\label{sec:obs}

\subsection{Target Selection}
\label{sec:target}
\noindent

We first obtained low-resolution ($R \approx 700$) spectroscopy of
eight fields in Sculptor using the f/2 camera of the IMACS
spectrograph \citep{dbh+11} at the Magellan-Baade telescope at Las Campanas Observatory. Each 
IMACS field spans a diameter of 27.4\arcmin, and the eight fields together produce nearly
complete coverage of the upper three magnitudes of Sculptor's red giant
branch (RGB) over a $37\arcmin \times 39\arcmin$\ area centered on the
galaxy, which approximately corresponds to complete coverage out to $\sim2$ times the core radius of Sculptor \citep{b+07}.  
The IMACS observations were taken with a narrow-band Ca~K
filter attached to a 200-lines~mm$^{-1}$ grism.  With this setup,
approximately 900 stars can be observed at a time.  IMACS targets were
selected from the photometric catalog of \citet*{cdb+05} using a broad
window surrounding the RGB so as not to exclude stars at the extremes
of the metallicity distribution.  The selection limits were based on a
Padova isochrone \citep{mgb+08} passing through the Sculptor RGB, and
extended from 0.37~mag bluer than the isochrone to 0.19~mag redder
than the isochrone in $V-I$, down to $V = 20$.

We selected Sculptor stars from the IMACS spectra for more extensive spectroscopic follow-up
observations. We identified a sample of
low-metallicity candidates by searching for stars with the smallest
Ca~K equivalent widths, adjusting for the color of each star according
to the calibration of \citet{brn+99}.  The most metal-poor known
Sculptor stars from \citet{fks+10} and \citet{tjh+10} were
independently recovered in this data set, as well as two new [Fe/H] $< -3.5$ stars \citep{sjf+15}. 
We then obtained $R \sim 4000$ and $R \sim 6000$
optical spectra of 22 of the best candidates,
using the MagE spectrograph \citep{mbt+08} at the Magellan telescopes. The majority of the
observed stars were confirmed as EMP stars, including a
number with spectra dominated by carbon features.


\begin{figure*}[!htbp]
\centering
\includegraphics[width =0.49\textwidth]{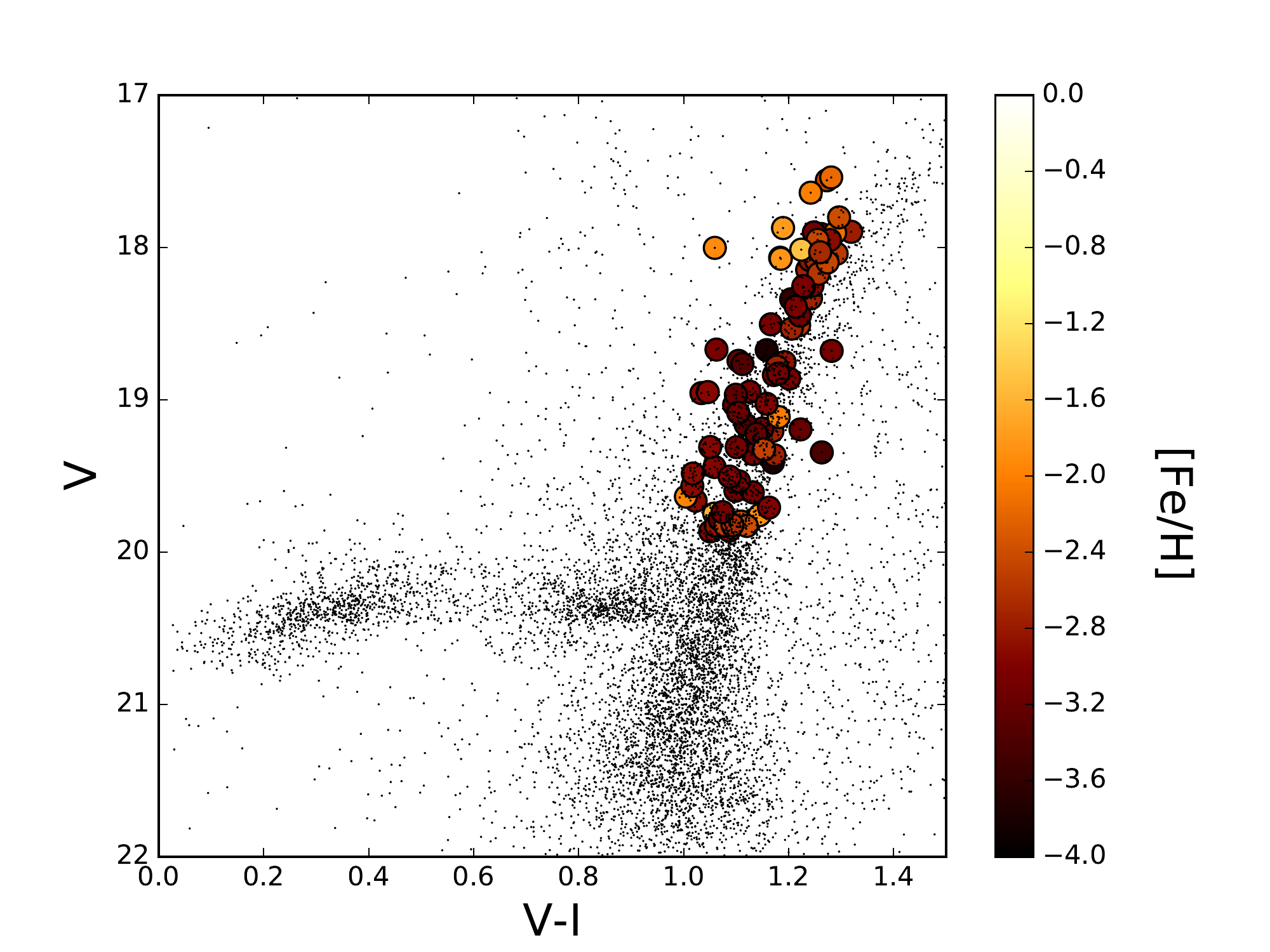}
\includegraphics[width =0.49\textwidth]{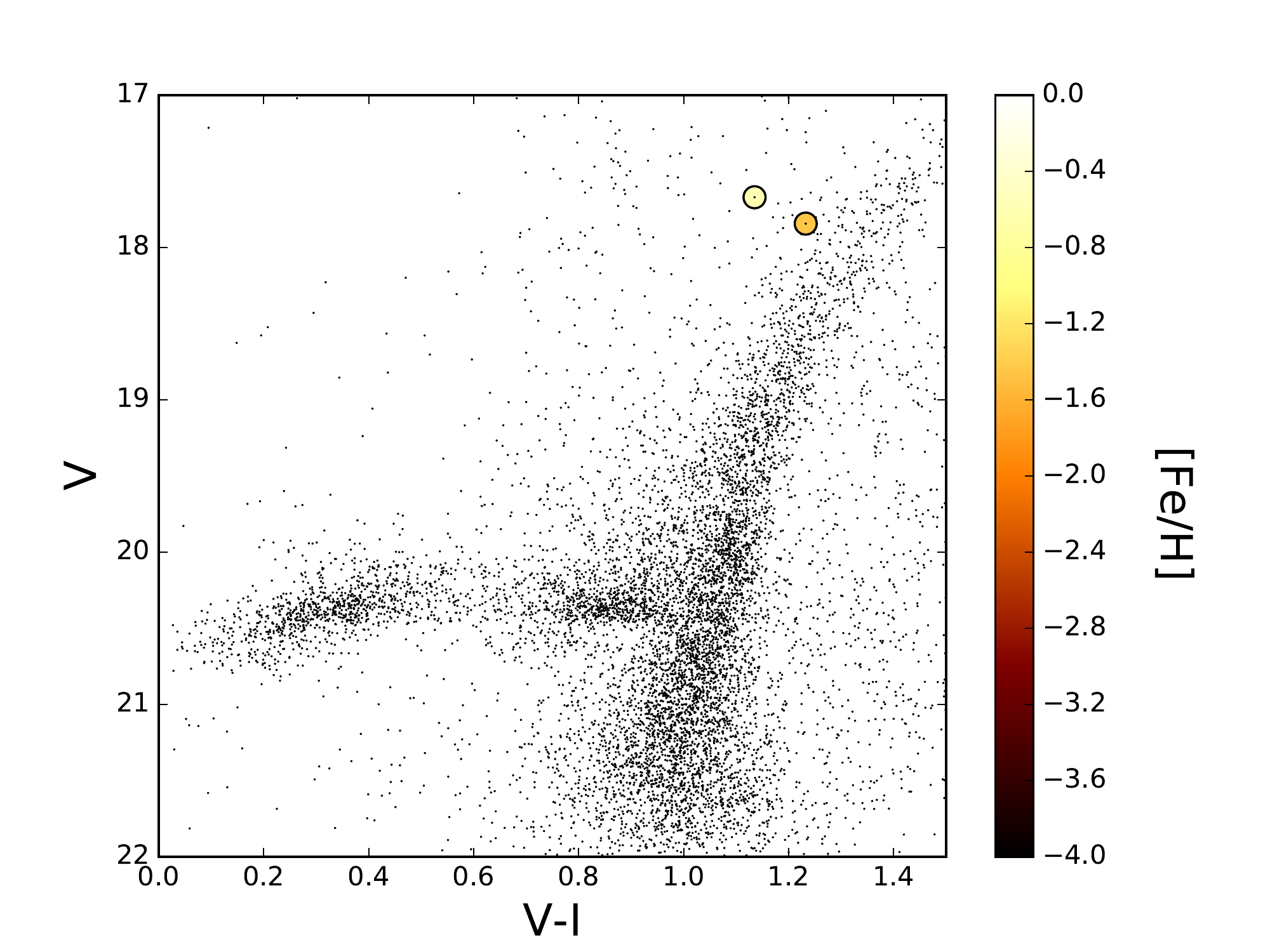}
\includegraphics[width =0.49\textwidth]{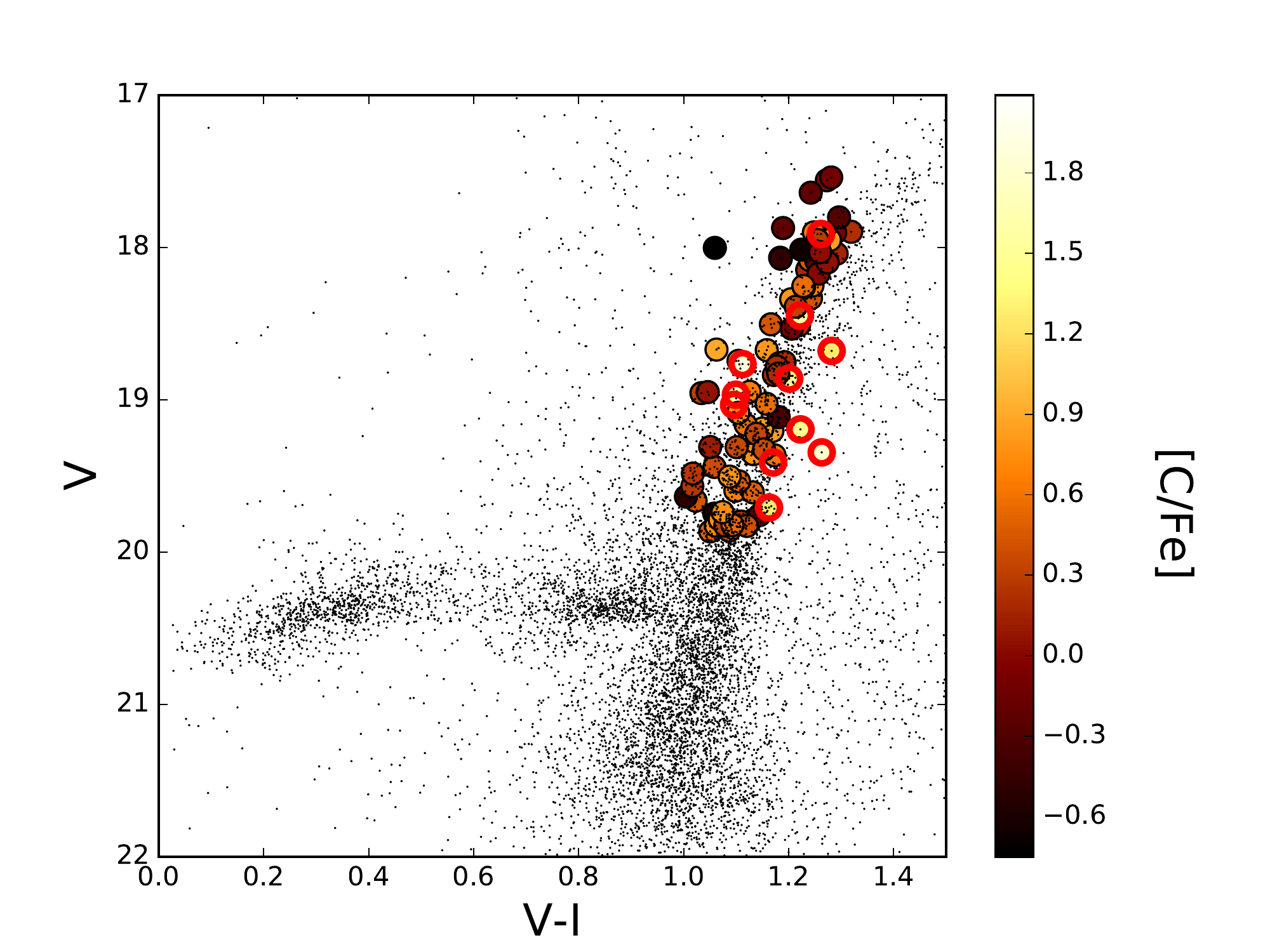}
\includegraphics[width =0.49\textwidth]{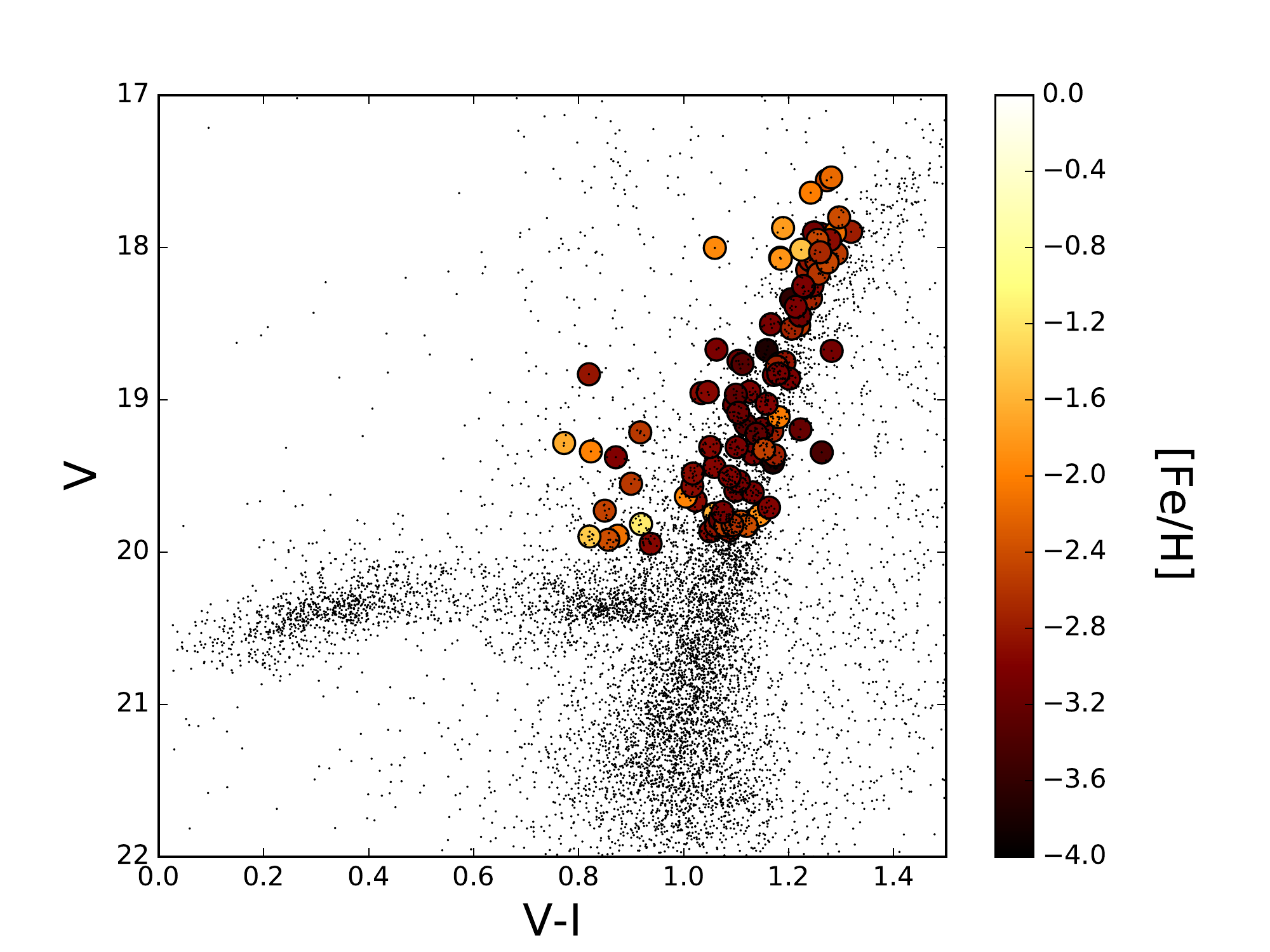}
\caption{Color magnitude diagrams (CMDs) of Sculptor from \citet{cdb+05}. M2FS targets
for which [Fe/H] and [C/Fe] are computed are over plotted. Top left: [Fe/H] of stars on the red giant branch of Sculptor
that were selected as the most metal-poor candidates. Top right: [Fe/H] of bright stars that were selected to fill available fibers. Much of the bright star sample was excluded from this work (see Section~\ref{sec:KP}). Bottom left: 
[C/Fe] of stars on the red giant branch of Sculptor that were selected to be metal-poor. Stars with saturated G-bands are circled in red. Bottom right:
[Fe/H] of all stars we observed that were selected to be metal-poor.}
\label{fig:CMD}
\end{figure*}


\subsection{M2FS Observations}

Having confirmed the utility of the IMACS data for both identifying EMP
and carbon-rich candidates in Sculptor, we set out to obtain medium-resolution spectra of a much
larger number of EMP candidates.  We observed two partially
overlapping 29.5\arcmin-diameter fields in Sculptor using the
Michigan/Magellan Fiber System (M2FS) \citep{mbc12} on
the Magellan--Clay telescope.  We
employed the low-resolution mode of M2FS, producing $R=2000$ spectra
covering $3700-5700$~\AA\ for 256 fibers.

M2FS targets were selected in two categories.  First, we chose all of
the EMP candidates from the IMACS sample (including those confirmed as
low metallicity and/or carbon-rich with MagE spectra).  Since these
candidates only occupied about half of the M2FS fibers, we then added
a magnitude-limited ``bright'' sample containing all stars along the Sculptor RGB 
brighter than $V=18.1$ in field 1 and $V=18.0$ in field 2
(the difference between the two reflects the number of fibers
available and the number of bright stars in each field).  This bright
sample should be unbiased with respect to metallicity or carbon
abundance.  About 30 fibers per field were devoted to blank sky
positions. A few broken fibers were not used.  The first M2FS
field, centered at RA (J2000), Dec (J2000) = 00:59:26, $-33$:45:19, was
observed for $5\times900$~s on the night of 23 November, 2013.  The second
M2FS field, centered at 01:00:47, $-33$:48:39 was observed for a
total of 6838\,s on 14 September, 2014. Figure~\ref{fig:CMD} shows
the M2FS targets for which [Fe/H] and [C/Fe] were measured in this work on
color magnitude diagrams of Sculptor. 
We note that stars with saturated CH G-bands are circled in red in the bottom left panel of Figure~\ref{fig:CMD}.
While the most carbon-enhanced stars do appear to be biased redward of the Sculptor RGB, they are not excluded from our selection procedure.

M2FS data were reduced using standard reduction techniques \citep{obg+16}. 
We first bias-subtracted each of the four amplifiers and merged the data.
We then extracted 2D spectra of all the fibers by using the spectroscopic flats to trace the location of 
science spectra on the CCD, flattened the science data, 
and took the inverse variance weighted average along the cross-dispersion axis
 of each science spectrum to extract a 1D spectrum.

We computed wavelength solutions using spectra of HgArNeXe and ThAr
calibration arc lamps. The typical dispersion of our wavelength solution was $\sim0.10$\,\AA, which
we derived by fitting third-degree polynomials to the calibration lamp spectra for the 2013 data. We derived the wavelength solution for the 2014 data by fitting third-degree Legendre polynomials.
We performed the sky-subtraction by fitting a fourth-order b-spline to the spectra of $\sim10$ sky fibers 
on the CCD, and fitting a third-order polynomial to the dependence of these spectra on the cross-dispersion direction
of the CCD (e.g., the location of the fiber's output on the CCD). We then subtracted the predicted sky model 
at the location of each science spectrum on the CCD, and extracted final 1D spectra.

\subsection{Follow-up MagE Observations}
\label{sec:MagE}
Motivated by the number of EMP and CEMP candidates from the M2FS data, 
we observed an additional ten Sculptor stars using the MagE spectrograph on the Magellan-Baade telescope in September 2016. 
This brought the total number of Sculptor stars observed with MagE to 31 stars, as one star had already been observed as part 
of the original 22 star sample (see Section~\ref{sec:target}).
Five of these ten stars showed strong carbon features in their M2FS spectra.
Another five were not seen to be as carbon-enhanced
from their M2FS spectra, but we chose to observe them due to their similar stellar parameters to the strongly carbon-enhanced stars.
These ten stars were analyzed to corroborate our M2FS carbon measurements.
We also observed the halo CEMP-r/s star CS29497-034 for reference purposes.
Five stars (four CEMP candidates and CS29497-034) were observed with the $0.7\arcsec$ slit ($R\sim6000$), which 
granted sufficient resolution to resolve barium lines at 4554\,\AA, 4934\,\AA, 5853\,\AA, 6141\,\AA, \text{and} 6496\,\AA. 
The remaining stars were observed with the $1.0\arcsec$ slit ($R\sim4000$). The MagE spectra were reduced using the Carnegie Python
pipeline described by \citet{k+03}. With these observations, we confirmed the CEMP and regular metal-poor nature of our candidates,
as suggested by the M2FS observations.


\begin{figure*}[!htbp]
\centering
\includegraphics[width = 0.9\textwidth]{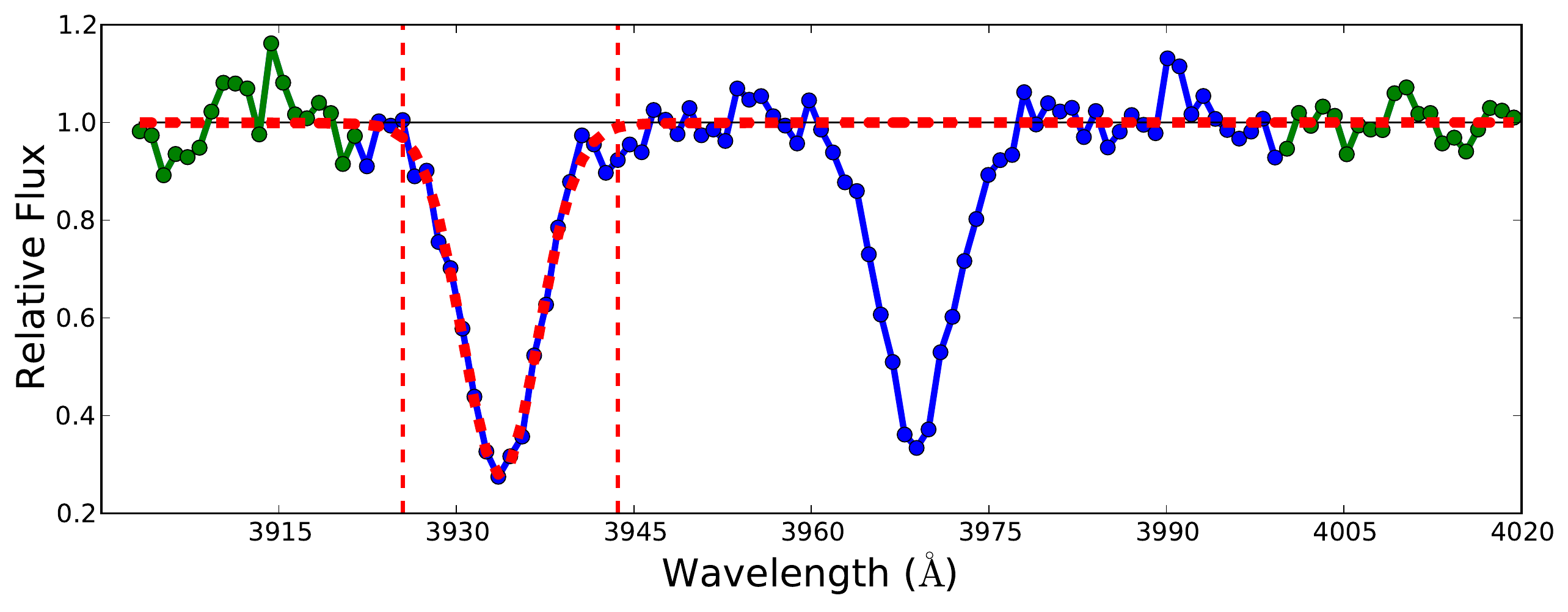}
\caption{Spectral region around the Ca II K line (3933.7\,{\AA}) after continuum normalization. The horizontal black dashed line depicts the continuum fit to the blue and 
red sidebands (green), and the vertical red dashed lines correspond to 
the range of integration for the KP index. The over-plotted dashed red line corresponds to the best fit Voigt profile. }
\label{fig:KPexample}
\end{figure*}


\section{Metallicity Measurements}
\label{sec:Metallicity}

We used established calibrations of two spectral line indices to measure [Fe/H] from the M2FS spectra. The first such index is the 
KP index, a measure of the equivalent width of the Ca II K line at 3933.7\,\AA. The second index is
the LACF index, a line index derived from applying the autocorrelation 
function (ACF) to the wavelength range 4000\,\AA\, to 4285\,\AA, which is chosen due to the presence
of many weak metal lines. Both line indices, along with the nature of their calibration to
[Fe/H] values, are thoroughly discussed by \citet{brn+99} and their implementation
in this work is detailed in this subsection.

\subsection{Membership Selection}
\label{sec:membership}
We measured radial velocities for each star primarily to exclude non-members of Sculptor.
Radial velocities were measured by cross-correlating the spectrum of each star with a rest-frame spectrum of the metal-poor giant HD122563.
Wavelength calibration for spectra obtained in 2013 was carried out using a ThAr lamp, resulting in a well calibrated range from 3900\,\AA\,\,to 5500\,\AA.
For the cross-correlation, we used this full range to determine velocities.
However, spectra obtained in 2014 had associated HgArNeXe arc lamp frames taken, which provided fewer usable reference lines.
 It was found that cross-correlating over only the H$\beta$ line (4830\,\AA\,\,to 4890\,\AA) gave the most precise ($\sim$10\,km\,s$^{-1}$) velocity measurements for these spectra.
Moreover, velocities obtained from the M2FS fiber observations in 2014 had to be adjusted to ensure that the mean velocity of the stars was centered on the velocity of Sculptor. 
Accordingly, velocities measured based on fiber observations on the red CCD chip were increased by 35\,km\,s$^{-1}$. Those from the blue CCD chip observations were increased by 31\,km\,s$^{-1}$.
For stars on both the 2013 and 2014 fiber plates, we used the velocity measurement from the 2013 spectrum.

We assumed that stars with velocities within 35\,km\,s$^{-1}$ of the systemic velocity of Sculptor were members.
This threshold corresponded to roughly 2.5$\sigma$ of our distribution of velocities after excluding outliers.
We found that applying this membership criterion recovered known members of Sculptor from \citet{wmo+09} and \citet{kgb+09}.
Using this criterion, we excluded four stars in our sample that would otherwise have been part of this data set.

\subsection{Stellar Parameters}
\label{sec:atm}
We derive initial $B-V$ color, $T_{\text{eff}}$, and $\log g$ estimates of stars in our IMACS sample by 
transforming $V$ and $I$ band photometry from \citet{cdb+05} using a 12 Gyr, [Fe/H] = $-$2.0 
Dartmouth isochrone \citep{dcj+08}. 
After a first pass measurement of [Fe/H] with this initial $B-V$ 
estimate (see Section~\ref{sec:Metallicity}), we iteratively update the metallicity 
of the isochrone and re-derive parameters until convergence. 
Before any measurement of [Fe/H], the spectrum was shifted so that the Ca II K line was centered at 3933.7\,\AA.
This re-centering was necessary given that the wavelength calibration was not necessarily accurate around the Ca II K feature, since only there was only one line below 4000\,\AA\,\,(a weak Ar II line at 3868.53\,\AA) in our arc frames.

\subsection{KP Index}
\label{sec:KP}
The KP index is a measurement of the pseudo-equivalent width of the Ca II K line at 3933.7\,\AA. 
To determine final KP indices, we first compute the K6, K12, and K18 indices
using bandwidths of $\Delta\lambda$ = 6\,\AA, 12\,\AA, and 18\,\AA, respectively, 
when calculating the equivalent width of the Ca II K feature \citep{bkg+90}. 
Table~\ref{tab:KP} lists the bands of these indices.
The KP index assumes the value of the K6 index when K6 $< 2$\,\AA,
the K12 index when K6 $> 2$\,\AA\, and K12 $< 5$\,\AA, and the K18 index 
when K12 $>5$\,\AA. 

To derive an estimate of the local continuum around the Ca II K feature, we 
fit a line through the red and blue sidebands listed in Table~\ref{tab:KP}.
We then visually inspected each continuum placement and 
applied a manual correction for a small subset of our sample that had 
an obviously bad fit (e.g., due to low S/N or nearby absorption features).
After continuum normalization, we derived estimates of the K6, K12, and K18 indices
using two methods. For the first approach,  we directly integrated across the line band to estimate the 
pseudo-equivalent width. For the second approach, we integrated over the best-fit Voigt profile to the Ca II K line as 
illustrated in Figure~\ref{fig:KPexample}. These two methods gave largely similar results, but the 
KP values from direct integration were adopted to ensure consistency with previous work involving the calibration. 
We derive [Fe/H] values using the KP index and $B-V$ color as inputs to the \citet{brn+99} calibration.

The KP index calibration from \citet{brn+99} is only valid for stars with $B-V$ $\le$ 1.2, meaning it can only be
readily applied to 100 stars in our sample. This population largely excludes the bright-star
sample, which is unbiased with respect to metallicity.


\begin{deluxetable}{cccc} 
\tablecolumns{4}
\tablewidth{\columnwidth}
\tablecaption{KP line indices (\AA)}
\tablehead{   
  \colhead{Line} &
  \colhead{Blue} &
  \colhead{Red} &
  \colhead{Band} \\
   \colhead{Index}&
   \colhead{Sideband}&
   \colhead{Sideband}
  &
}
\startdata
K6   & 3903$-$3923 & 4000$-$4020 & 3930.7$-$3936.7\\  
K12 & 3903$-$3923 & 4000$-$4020  & 3927.7$-$3939.7\\
K18 & 3903$-$3923 & 4000$-$4020  & 3924.7$-$3942.7
\enddata
\label{tab:KP}
\end{deluxetable}



\begin{figure*}[!htbp]
\centering
\includegraphics[width =1.0\textwidth]{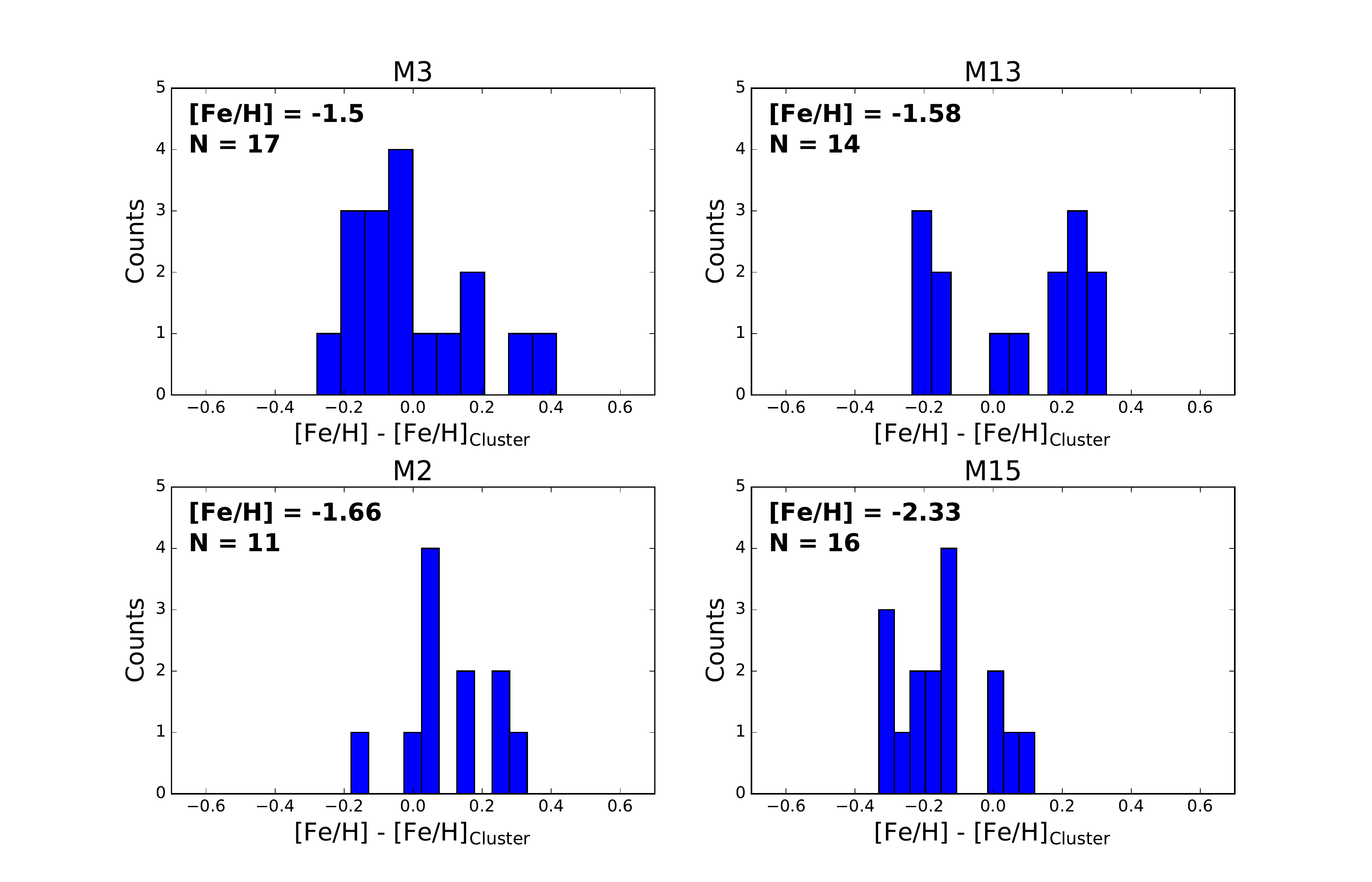}
\caption{Histograms of the difference between our measured metallicity of each globular cluster member and the overall cluster metallicity for globular clusters M3 (top left), M13 (top right), M2 (bottom left), and M15 (bottom right).}
\label{fig:GlobSeparate}
\end{figure*}


\subsection{LACF Index}

The LACF index measures the strength of many weak metal lines between 
4000\,\AA\, and 4285\,\AA\,\citep{rf+89,brn+99}. It is computed by taking the autocorrelation 
of a spectrum within the aforementioned wavelength range after excising extraneous line features.
The LACF index is then defined as the $\log$ of the value of the autocorrelation function (ACF) 
at $\tau = 0$ as defined in Equation~\ref{eqn:ACF} over this interval.

To ensure we computed the LACF index in a manner consistent with \citet{brn+99}, we 
closely reproduced their methodology. We
first interpolated each spectrum using a cubic spline and re-binned
in 0.5\,\AA\,increments to match their calibration sample.
We then excised the ranges 4091.8\,\AA\, to 4111.8\,\AA\, and 4166\,\AA\, to 4216\,\AA\, 
to remove effects from the H$\delta$ region and CN molecular absorption, respectively. To calculate the continuum,
each of the three resulting ranges were independently fit by a fourth-order polynomial, 
after which outliers 2$\sigma$ above and 0.3$\sigma$ below each fit were excluded. 
An acceptable continuum estimate was returned after four iterations of this
process. 

After normalizing each wavelength segment by its corresponding continuum 
estimate, we re-stitched the three segments together and computed the power spectrum
of the resulting spectrum. We then set the high and low frequency components 
of the power spectrum to zero in order to remove the effects of high frequency 
noise and continuum effects, respectively. The inverse Fourier 
transform of the power spectrum was taken to derive the ACF, which was 
then divided by the square of the mean counts in the normalized region. 
We finally computed the LACF index by taking the $\log$ of the resulting ACF at $\tau = 0$.

It is important to note that an alternative expression of the autocorrelation function is 
\begin{equation}
\text{ACF}(\tau) = \int_{\nu_1}^{\nu_2} f(\lambda + \tau) \bar{f}(\lambda)d\lambda
\label{eqn:ACF}
\end{equation}
where $\bar{f}(\lambda)$ is the complex conjugate of the function $f(\lambda)$.
From Equation~\ref{eqn:ACF}, it is clear that computing the LACF index, defined
as the $\log$ of the value of the ACF at $\tau = 0$,
is analogous to integrating the squared spectrum after manipulating 
Fourier components to remove continuum and noise related effects.
This fact motivates the application of an ACF to measure line strength.
As with the KP index, the LACF index is only calibrated to [Fe/H] for stars 
with $B-V \le 1.2$ (see discussion in Section~\ref{sec:KP}).

\subsection{Comparison of Methods and Final [Fe/H] Values}

To ensure our measured KP and LACF indices were consistent with the existing [Fe/H] calibration, 
we measured KP and LACF indices on a subset of the calibration sample in 
\cite{brn+99}. We found agreement in KP indices, 
but a gradually increasing scatter in LACF measurements when
LACF $<$ 0, which roughly corresponds to very 
metal-poor stars, stars with high effective temperatures, or stars with 
spectra that have low signal-to-noise. We thus chose to discard the LACF-based metallicity 
measurement for stars with LACF $< -0.5$ or when [Fe/H]$_{\text{KP}} < -2.5$. 
Since the LACF works best at measuring [Fe/H] in the more metal-rich
regime where weak metal lines are more prominent, this exclusion seems reasonable.
We also chose to discard KP-based metallicity measurements when [Fe/H]$_{\text{KP}} > -1.0$,
motivated by the failure of the KP calibration at high metallicities due to the saturation of the 
Ca II K line. In the regime where both KP and LACF based metallicities are valid, we take the average of
the two measurements weighted by the measurement uncertainty.

The $\alpha$-element abundance of stars in the \citet{brn+99} calibration
is assumed to be [$\alpha$/Fe] $ = +0.4$ for [Fe/H] $< -1.5$ and 
[$\alpha$/Fe] $= -0.27 \times $[Fe/H] for $-1.5 <$ [Fe/H] $< 0$.
Stars in Sculptor display a different trend in [$\alpha$/Fe] with [Fe/H].
We account for this discrepancy by first computing an [$\alpha$/H] measurement
for our stars based on the aforementioned $\alpha$-element trends used in the 
Beers calibration for both the
KP and LACF derived metallicities.
We then fit a line to a Sculptor [Fe/H] vs. [$\alpha$/H] trend derived from measurements in \citet{kgb+09},
and use this trend to compute an [Fe/H] measurement from our [$\alpha$/H] measurement
for each of our Sculptor stars. This adjustment is motivated by the fact that
the \citet{brn+99} calibrations measure the strength of $\alpha$-element features and 
derive metallicities under the assumption of a given [$\alpha$/Fe] for halo stars, which is discrepant from
the trend in dwarf galaxy stars. 
This correction typically increased the metallicities of stars in our sample by $\lesssim$\,0.1\,dex, since it had no effect on stars with [Fe/H] $< -3.0$ and increased metallicities of stars with [Fe/H] = $-2.5$ by $\sim$\,0.1\,dex.

Initial [Fe/H] uncertainties were assigned following \citet{brn+99}. 
To account for uncertainties in using an isochrone to transform 
between $V-I$ and $B-V$ color,
we propagated the uncertainty in our original $V-I$ color to the final
[Fe/H] measurements and added this effect in quadrature to the other uncertainties.
We also propagated uncertainties in the age of the isochrone, 
which had negligible effects. 
Finally, we re-measured the metallicities after shifting the continuum by the standard errors of the fluxes in the red and blue continuum regions.
The difference between the re-measured metallicities and the original metallities was added in quadrature with the other estimates of uncertainty.
Typical uncertainties are $\sim$0.25\,dex.


\begin{figure}[!htbp]
\centering
\includegraphics[width = \columnwidth]{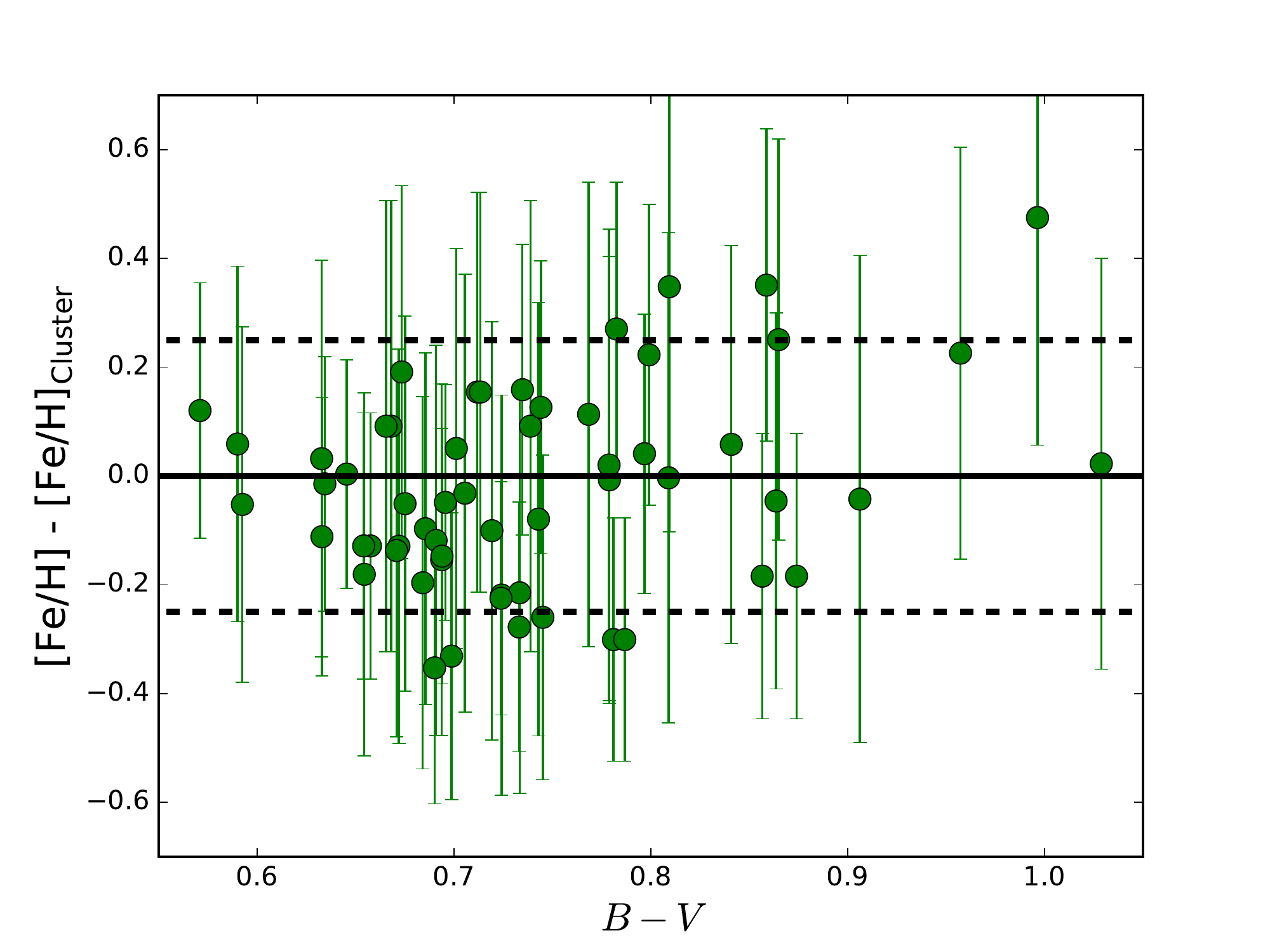}
\caption{Difference between our measured metallicity of each cluster member and the overall cluster metallicity as a function of
$B-V$ color. Dashed lines correspond to $\pm0.25\,$dex. The mean of the distribution of residuals is $-0.02$ and the standard deviation is $0.18$.}
\label{fig:GlobCombined}
\end{figure}



\begin{deluxetable*}{ccccccccc} 
\tablecolumns{8}
\tablewidth{\textwidth}
\tablecaption{Stellar Parameter Comparison}
\tablehead{   
   \colhead{ID} &
   \colhead{$\log{g}_{\text{MR}}$}&
   \colhead{$\log{g}_{\text{HR}}$}&  
   \colhead{T$_{\text{eff}_\text{MR}}$(K)}&
   \colhead{T$_{\text{eff}_\text{HR}}$(K)}& 
   \colhead{[Fe/H]$_{\text{KP}}$} &
   \colhead{[Fe/H]$_{\text{HR}}$} &
   \colhead{$\Delta$[Fe/H]} &
   \colhead{Ref.} 
}
\startdata
S1020549 & 1.30 & 1.25 & 4610 & 4702 & $-3.74\pm0.21$ & $-$3.68 & $-0.06$ & S15\\
Scl 6\underline{\hspace{0.2cm}}6\underline{\hspace{0.2cm}}402 & 1.67 & 2.00 & 4796 & 4945 & $-3.91\pm0.25$ &$-$3.53 & $-0.38$ & S15\\
Scl 11\underline{\hspace{0.2cm}}1\underline{\hspace{0.2cm}}4296 & 1.52 & 1.45 & 4716 & 4770 &  $-3.90\pm0.21$& $-$3.77 & $-0.13$ & S15\\
Scl 07$-$50 & 1.35 & 1.05 & 4676 & 4558 &  $-3.96\pm0.20$& $-$4.05 & $+0.09$ & S15\\
\noalign{\vskip 0.8mm} 
\hline
\noalign{\vskip 1.4mm} 
S1020549 & 1.29 & 1.25 & 4581 & 4702 & $-3.63\pm0.21$ & $-$3.68 & $+0.05$ & S15\\
Scl 11\underline{\hspace{0.2cm}}1\underline{\hspace{0.2cm}}4296 & 1.55 & 1.45 & 4697 & 4770 &  $-3.33\pm0.22$& $-$3.77 & $+0.44$ & S15\\
Scl 07$-$50 & 1.40 & 1.05 & 4641 & 4558 & $-3.77\pm0.20$ & $-$4.05 & $+0.28$ & S15\\
ET0381 & 1.19 & 1.17 & 4532 & 4540 & $-2.83\pm0.19$ &$-$2.83 & $+0.00$ & J15\\ 
Scl\underline{\hspace{0.2cm}}03\underline{\hspace{0.2cm}}059 & 1.10 & 1.10 & 4492 & 4400 & $-3.00\pm0.15$ & $-$3.20 & $+0.20$ & J15

\enddata
\tablecomments{[Fe/H]$_{\text{KP}}$ is the metallicity measured by applying the KP index 
calibration. [Fe/H]$_{\text{HR}}$ is the metallicity measured in the indicated reference paper. 
Measurements labeled MR are medium-resolution measurements following the methodology of this paper. 
Top section: Measurements from smoothed high-resolution spectra of stars presented in \citet{sjf+15}.
Bottom section: Measurements from our medium-resolution M2FS spectra.
S15 and J15 refer to \citet{sjf+15} and \citet{jnm+15},
respectively. Log $g$ values in this table
have been corrected by +0.39\,dex to account for the measured offset with respect to \citet{kgs+10}.}
\label{tab:KPcompare}
\end{deluxetable*}


\begin{figure*}[!htbp]
\centering
\includegraphics[width = \columnwidth]{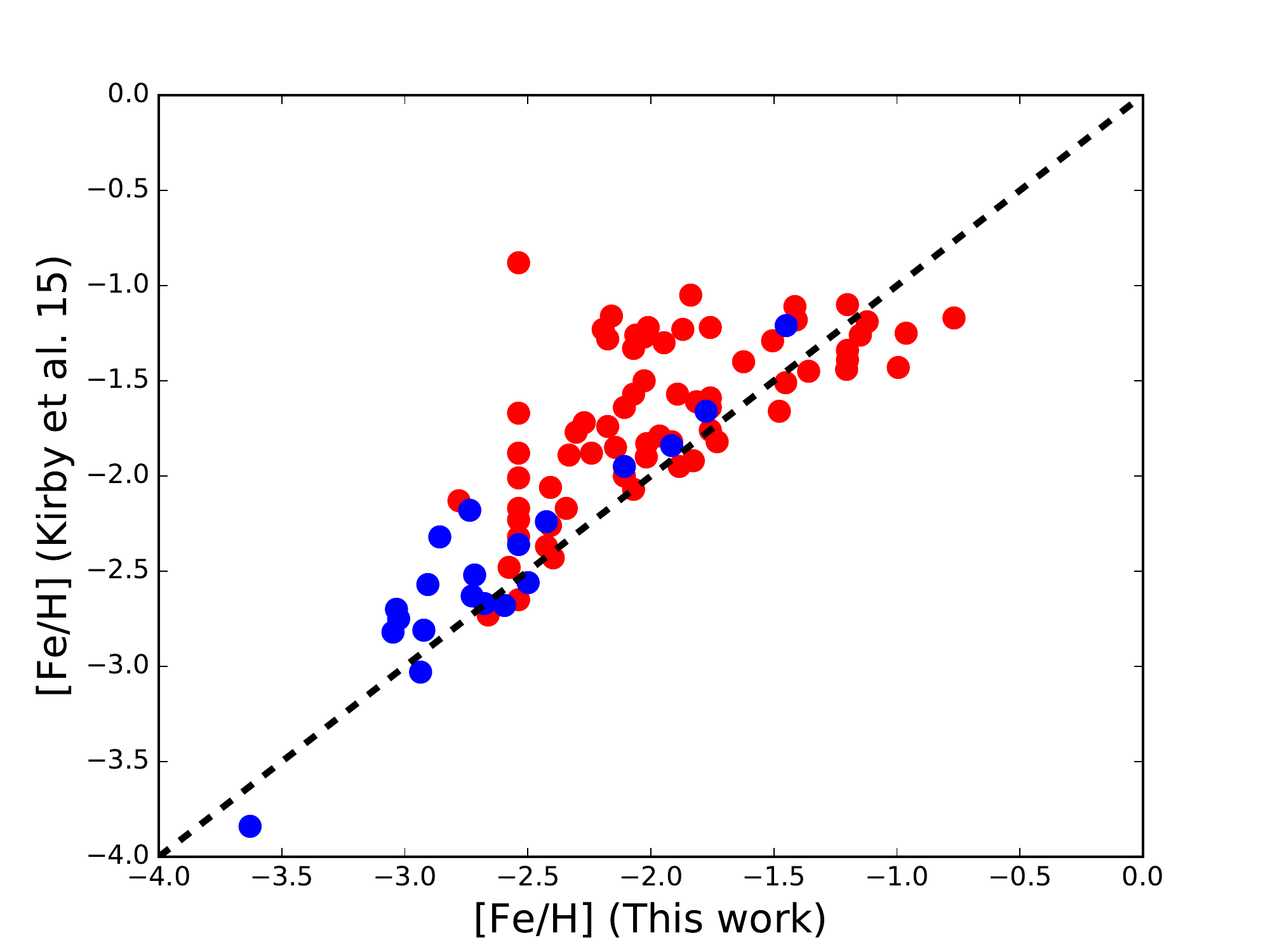}
\includegraphics[width = \columnwidth]{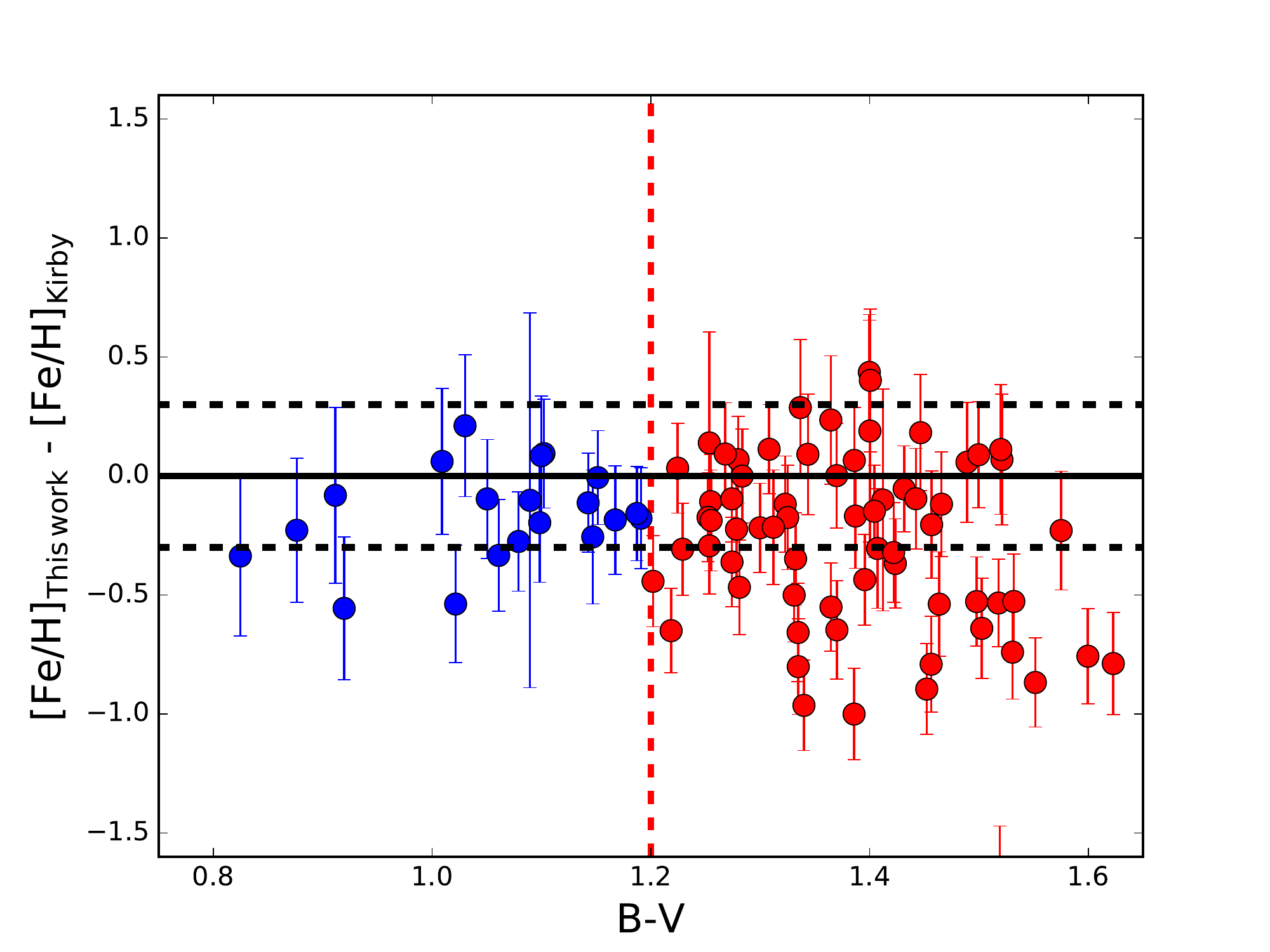}
\caption{Left: Comparison of [Fe/H] measured by \citet{kgs+10} and [Fe/H] measured in this work for the 86 stars in both samples. Blue points correspond to stars with $B-V \le 1.2$ and red points correspond to stars with $B-V > 1.2$. Right: The difference between [Fe/H] measured in this work and [Fe/H] measured by \citet{kgz+15} as a function of $B-V$ color. The vertical line marks the cutoff to the right of which $B-V$ colors are not directly calibrated to [Fe/H] in \citet{brn+99}. Dashed lines indicate $\pm0.30\,$dex.}
\label{fig:K10Comp}
\end{figure*}



\begin{figure}[!htbp]
\centering
\includegraphics[width = \columnwidth]{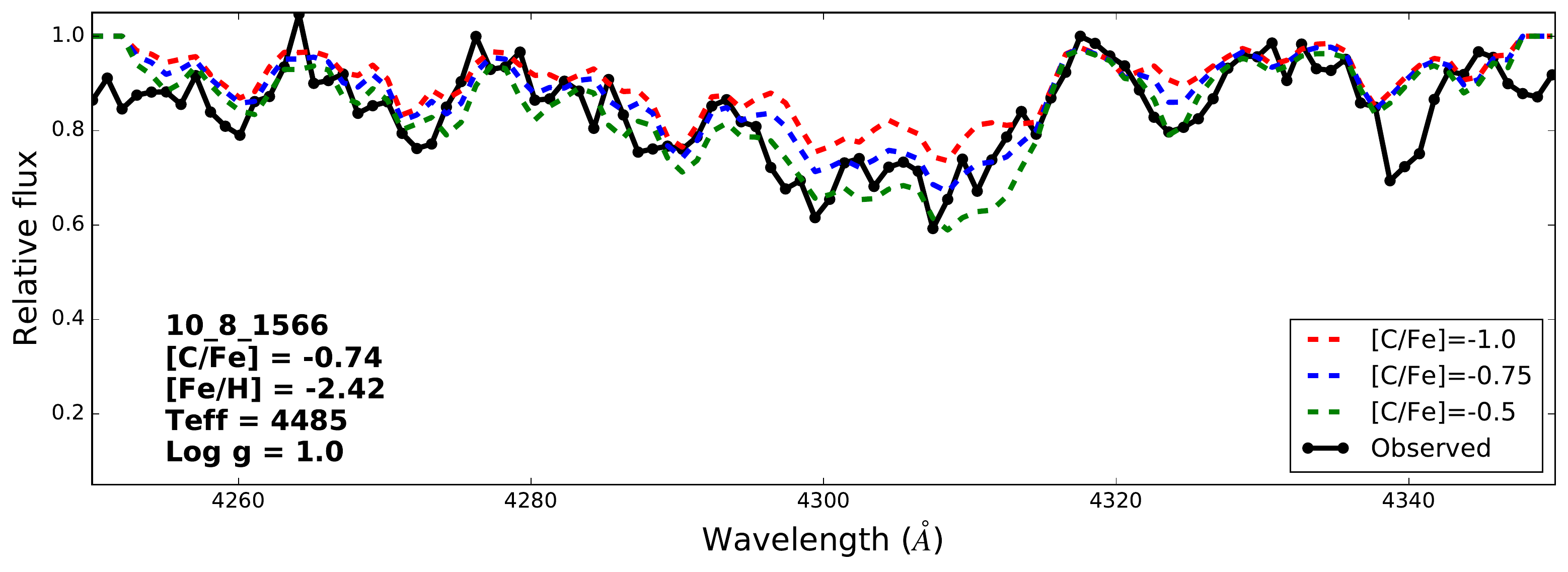}
\includegraphics[width = \columnwidth]{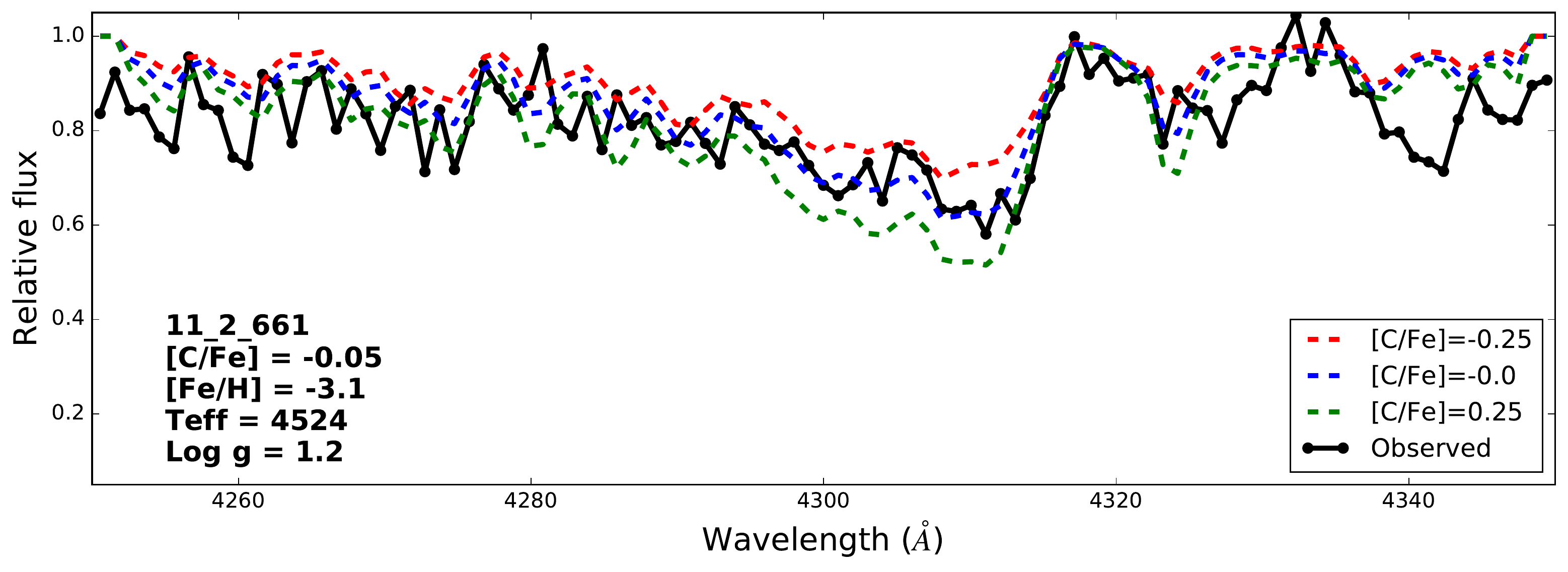}
\includegraphics[width = \columnwidth]{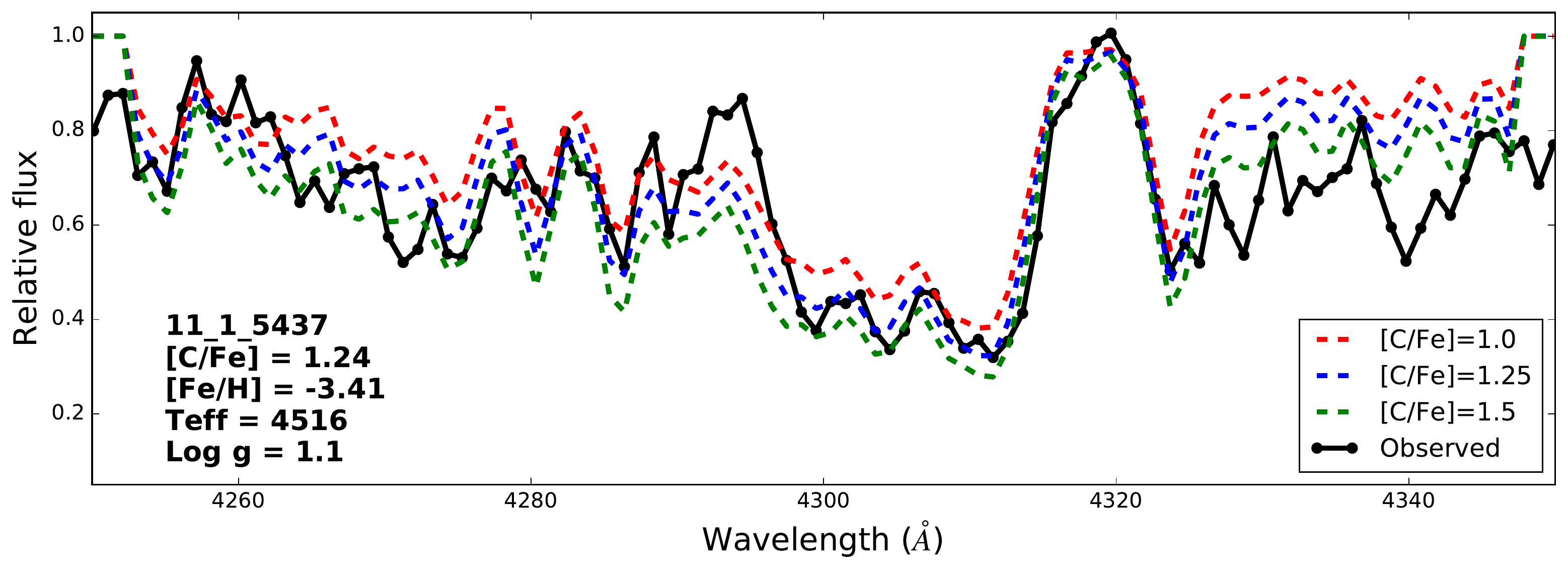}
\caption{Spectral region around the G-band together with best fitting synthetic spectra (blue) for three example observed M2FS spectra (black). Synthetic spectra with [C/Fe] closest to the $1\sigma$ upper and lower
[C/Fe] measurements are overplotted in red and green, respectively.}
\vspace{0.50cm}
\label{fig:SynthFits}
\end{figure}


\subsection{External Validation: Comparison to Globular Cluster Members}

As an external check on our metallicity measurements,
we determined [Fe/H] values for cool ($T_{\text{eff}} < $ 5500\,K) member stars in four 
globular clusters (M2, M3, M13, M15) with metallicities ranging from [Fe/H] = $-$2.33 to $-$1.5.
We retrieved medium-resolution spectra of these stars from
the Sloan Digital Sky Survey-III\footnote[1]{\texttt{http://dr10.sdss3.org}} \citep{ewa+11, aaa+14}. 
$V-I$ colors were derived by applying an empirical 
color transformation following \citet{jga+06}.

The metallicity spread among members of a globular cluster is a fraction of our measurement uncertainties,
with the exception of some anomalies in M2 \citep{yrg+14}. 
Thus, we used the offset of our [Fe/H] values of each cluster member 
from the average metallicity of the globular cluster to gauge the validity of our metallicity calibration.
Before measuring metallicities, we recorded the mean $[\alpha/\text{Fe}]$ of these
globular clusters from \citet{c+96}, \citet{kgs+08}, and \citet{yrg+14} and corrected them for the discrepant [$\alpha$/Fe]
assumption in our calibrations. As shown in Figures~\ref{fig:GlobSeparate} and \ref{fig:GlobCombined}, 
our measurements gave largely reasonable results, with an overall [Fe/H] offset of $-0.02\,$dex and 
scatter of $0.18\,$dex. This is consistent with our typical derived uncertainty in [Fe/H] of
$\sim$0.25\,dex.

\subsection{External Validation: Comparison to Kirby et al.}
\label{sec:Kirby13}

\citet{kgb+09,kgs+10,kcg+13} measured the metallicities and $\alpha$-abundances of a total of 391 
stars in Sculptor with medium-resolution spectroscopic data from the Deep Imaging Multi-Object Spectrometer 
on the Keck II telescope. We found 86 stars in common with our full sample 
of $\sim250$ stars, of which 20 stars have $B-V \le$ 1.2. 
We compare the stellar parameter measurements between 
all 86 stars.
We find reasonable agreement in our $T_{\text{eff}}$ measurements as demonstrated by a mean offset of $\overline{\Delta T_{\text{eff}}} = 25$\,K and a standard deviation of $\sigma(\Delta\text{Teff}) =137\text{K}$.
For $\log g$, we correct the significant offset of +0.39\,dex compared to the Kirby et al. sample.
The mean difference in $\log g$ after this correction is 0, with a standard deviation of 0.17\,dex. If we were to only consider stars with $B-V \le$ 1.2, then the standard deviation would be 0.23\,dex.
This correction also results in agreements with $\log g$ values of stars with high-resolution spectroscopic stellar parameters (see Table~\ref{tab:KPcompare}). We note that not applying this gravity correction would artificially increase the carbon abundance correction we apply to take into account the evolutionary state of the star (see \citealt{pfb+14}), and thus the number of CEMP stars in the sample.
We then compare our metallicities for the subset of stars with $B-V \le 1.2$. 
We find a mean offset of $\text{[Fe/H]} - \text{[Fe/H]}_{\text{K10}}\approx-0.11$\,dex 
with a standard deviation of $\sim0.15$\,dex (excluding two outliers below $B-V = 1.2$ for which we measure a lower metallicity by over $\sim0.5$\,dex, see Figure~\ref{fig:K10Comp}). Including these outliers changes the mean offset to $\text{[Fe/H]} - \text{[Fe/H]}_{\text{K10}}\approx-0.16$\,dex and increases the scatter to $\sim0.19$\,dex.

Both outliers (10\_8\_2730 and 10\_8\_2788) in Figure~\ref{fig:K10Comp} have low reported calcium abundances ([Ca/Fe]=$-0.23\pm0.30$ and [Ca/Fe] = 0.05$\pm0.39$) in \citet{kgs+10}. This could lead to a weaker Ca II K line than our assumed [$\alpha$/Fe] would suggest
and would cause an underestimation of the metallicity.

Figure~\ref{fig:K10Comp} also demonstrates the failure of the KP and ACF calibrations for $B-V >$ 1.2. Accordingly,
we choose to limit this work to the subset of stars in our sample with $B-V \le$ 1.2.

\subsection{External Validation: Comparison to High-Resolution [Fe/H]}
\label{sec:hiresfeh}
As a final check to ensure the KP calibration holds for extremely metal-poor (EMP) stars, 
we retrieved high-resolution spectra of four EMP Sculptor members from \citet{sjf+15}\footnote[2]{The spectrum of the 
fifth star in that paper does not extend blue-ward to the Ca II K feature.}. 
We smoothed these spectra to match the resolution of our medium-resolution data
and degraded the signal-to-noise ratio to 20\,\AA$^{-1}$.
We then computed KP-derived metallicities of these stars. 
The results are shown in the top
portion of Table~\ref{tab:KPcompare} and demonstrate the accuracy of KP calibration.

We also compared the KP-derived metallicities from our M2FS sample to high-resolution measurements in \citet{sjf+15}
and \citet{jnm+15} for five stars in common to both samples. The results are shown in the bottom panel of Table~\ref{tab:KPcompare}.
We note a marginally higher KP-derived metallicity in most cases for the EMP stars in the M2FS data.
The largest residual (11\_1\_4296) can reasonably be explained due to the presence of noise near the Ca II K line.
Interpolating over this noise spike results in a marginally lower disagreement of +0.34\,dex when compared to the high-resolution [Fe/H] measurement.

\section{Carbon Abundance Measurements}
\label{sec:carbon}

To derive carbon abundances ([C/Fe]), we matched each observed
spectrum to a grid of synthetic spectra closely following the methodology of \citet{kgz+15}. 
We generated these using the MOOG spectrum synthesis code with an updated treatment 
of scattering \citep{s+73,sks+11}, and model atmospheres from ATLAS9 \citep{ck+04}. 
We independently computed [C/Fe] using regression relations from \citet{rbs+05}, but found that 
fitting to a grid allowed accurate [C/Fe] measurements over a broader range of 
input parameters.

\subsection{Spectrum Synthesis}
\label{sec:synth}

Table~\ref{tab:synthgrid} lists the stellar parameters of the generated grid of synthetic spectra.
We used a comprehensive line list spanning 4100\,\AA\, to 4500\,\AA\, compiled by \citet{kgz+15}. 
The list comprises transitions from the 
Vienna Atomic Line Database \citep[VALD;][]{pkr+95, kpr+99}, the National 
Institutes of Standards and Technology \citep[NIST;][]{NIST_ASD}, \citet{k+92},
and \citet{jli+96}. We assumed an isotope ratio of $\ce{^{12}C}/\ce{^{13}C} = 6$
based on the low surface gravity ($\log g \le$ 2.0) of most of our stars. The $\alpha$-element
abundance of the grid was chosen to be +0.2\,dex, which is the mean expected
value for this sample of Sculptor members, as gleaned from measurements by
\citet{kgb+09}. Each synthetic spectrum was degraded to match the resolution 
of medium-resolution M2FS spectra.
This grid was then used for measuring the carbon abundances reported in this paper. 
It should have similar inputs (e.g., line lists, model atmospheres) to previous works on the CEMP fraction in dwarf galaxies \citep[e.g.,][]{kgz+15} and other studies of halo stars. 
This enables a fair comparison of our results with literature values.

To appropriately compare our [C/Fe] measurements with nearly all values in the literature,
we generated two smaller test grids based on model atmospheres and line lists different from those in the primary grid used in our analysis.
The first test grid was generated using the Turbospectrum synthesis code \citep{ap+98,p+12}, MARCS
model atmospheres \citep{gee+08}, and a line list comprised of atomic data from VALD, CH data from \citet{mpv+14}, 
and CN data from \citet{brw+14} and \citet{slr+14}. The second test grid had the same inputs
as the first test grid, but was generated using MOOG to compare differences
between just the two synthesis codes. Both test grids spanned 4500 to 4800\,K in effective
temperature, 1.0 to 2.0\,dex in log\,\textit{g}, and $-4.0$ to $-2.5\,$dex in [Fe/H], which roughly 
covers the stellar parameters of the more metal-poor stars in our sample.

\subsection{Fitting to the Grid}
\label{sec:gridfit}

Since synthetic spectra computed by MOOG are generated as normalized spectra,
we normalized each spectrum. We found that iteratively fitting a cubic spline 
to the observed data from 4100\,\AA\, to 4500\,\AA, excluding points 5$\sigma$ above and 0.1$\sigma$ below in each iteration,
reproduced the continuum well. After dividing the observed spectrum
by our continuum estimate, we found the best fitting synthetic spectrum by varying [C/Fe]. 

We then implemented a $\chi^2$ minimizer to match the region spanning the CH G-band 
(4260\,\AA\, to 4325\,\AA) to the synthetic grid. 
We measured [C/Fe] by setting the three parameters $T_\text{eff}$, log $g$, 
and [Fe/H] equal to the values determined from our medium-resolution M2FS measurements and letting 
[C/Fe] vary as a free parameter. We then interpolated between the five [C/Fe] measurements
around the best [C/Fe] value with the lowest $\chi^2$ values to determine a final carbon abundance. 
Sample fits are shown in Figure~\ref{fig:SynthFits}. Each [C/Fe] measurement was 
corrected to account for the depletion of carbon for stars on the upper red giant branch 
\citep{pfb+14}.
After this correction, we find no statistically significant trend in the [C/Fe] abundances with respect to measured Log\,\textit{g} values.

To determine the uncertainty in our carbon abundance measurements, we re-measured [C/Fe] 100 times for each spectrum after
varying the stellar parameters each time. For each measurement of [C/Fe], we drew values of T$_\text{eff}$, log $g$, 
[Fe/H] from gaussian distributions parametrized by the medium-resolution measurements and uncertainties of those parameters. 
We adopted stellar parameter uncertainties of $\pm150$\,K for $T_\text{eff}$ and $\pm$0.15\,dex for log $g$. 
Before each measurement, the continuum was multiplied by a number drawn
from a gaussian distribution centered on one with $\sigma = 0.01$ to capture the uncertainty in continuum placement. 
The standard deviation of the resulting [C/Fe] measurements was taken as the total uncertainty 
in our measurement.


\begin{deluxetable}{cccc} 
\tablecolumns{3}
\tablewidth{0.3\textwidth}
\tablecaption{Synthetic spectrum grid stellar parameter range}
\tablehead{   
  \colhead{Parameter} &
  \colhead{Minimum} &
  \colhead{Maximum} &
  \colhead{Step}
}
\startdata
$\lambda$ & 4250\,\AA & 4350\,\AA & 0.01\,\AA\\
$T_{\text{eff}}$ & 3700\,K & 5700\,K & 50\,K\\
log $g$ & 0.0 & 4.0 & 0.5\\
$[$Fe/H$]$ & $-$4.0 & $+$0.2 & 0.2\\
$[$C/Fe$]$ & $-$2.00 & 2.00 & 0.25\\
\label{tab:synthgrid}
\end{deluxetable}



\begin{figure}[!htbp]
\centering
\includegraphics[width = \columnwidth]{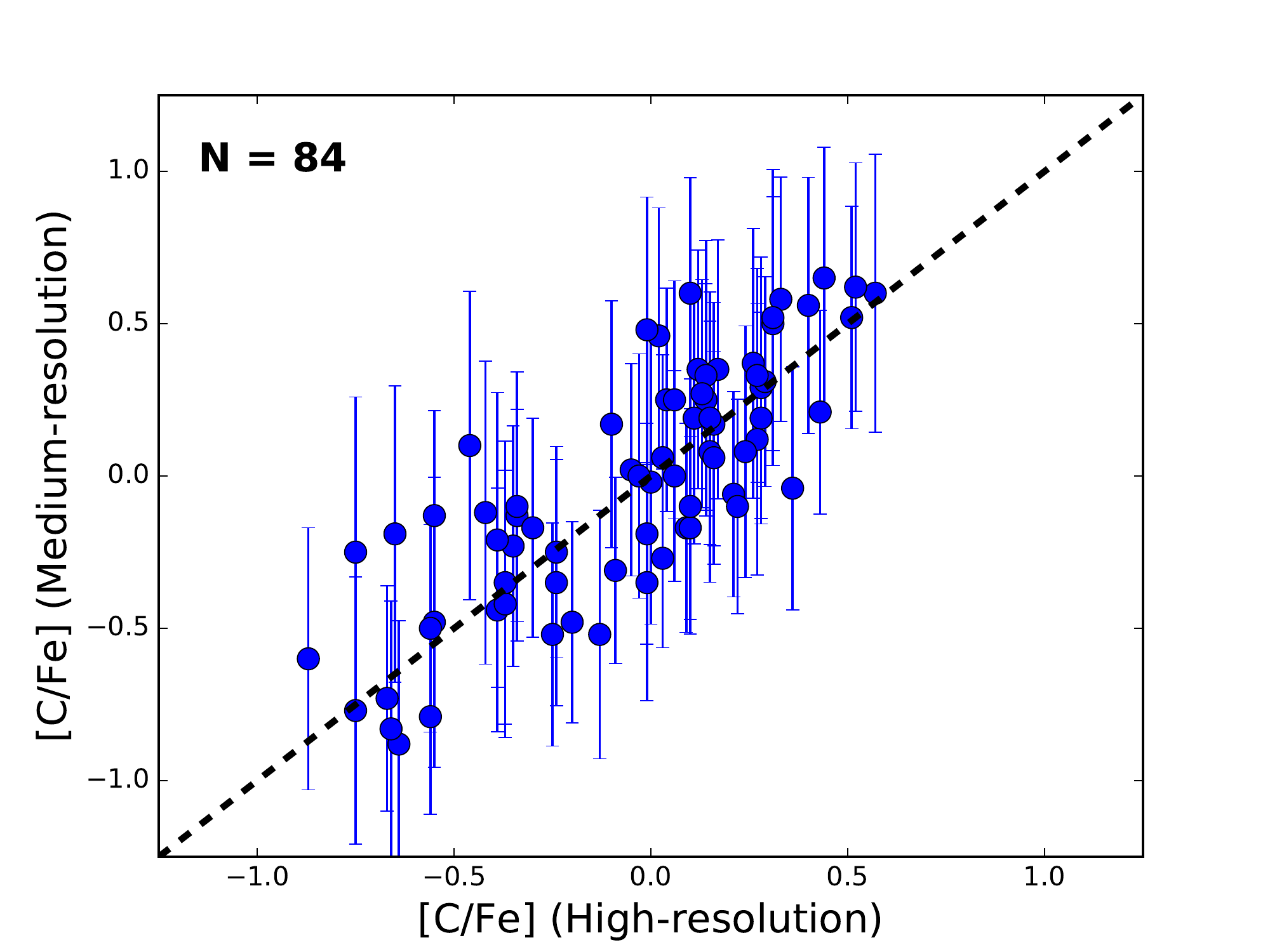}
\caption{Carbon abundance measurements of metal-poor stars from \citet{jkf+15} after spectra were degraded to the same resolution as the Sculptor M2FS spectra versus high-resolution [C/Fe] measurements of the same stars. The median offset between medium-resolution and high-resolution [C/Fe] measurements is $0.03$ dex and the observed scatter is $0.22$ dex.}
\label{fig:SkymapperTest}
\end{figure}

\subsubsection{External Validation: Comparison to SkyMapper Sample from \citet{jkf+15}}
\label{sec:carbonvalidation}

We applied our framework to measure [C/Fe] values to a 
sample of high-resolution Magellan/MIKE spectra of
metal-poor halo stars selected from the SkyMapper survey. These spectra were degraded 
to match the resolution of our medium-resolution spectra and were
injected with gaussian noise to bring the S/N down to 20\,\AA$^{-1}$. High-resolution [C/Fe]
abundances computed by \citet{jkf+15} were used as reference values.

Analyzing a sample of 84 stars,
we find that our [C/Fe] values differ from the high-resolution values [C/Fe] by a 
median value of 0.03\,dex with $\sigma(\Delta\text{[C/Fe]}) = 0.22$\,dex (see Figure~\ref{fig:SkymapperTest}). 
We regard this agreement as excellent, since different normalization routines tend to produce different [C/Fe] measurements,
given the difficulty of normalizing the G-band due to ubiquitous absorption features. 
Furthermore, the average offset is dwarfed by the typical measurement
uncertainty of $\sim0.35\,$dex. Raising the continuum placement by $2\%$ increases [C/Fe] by $\sim0.1\,$dex in this sample.

\begin{deluxetable}{cccccc} 
\tablecolumns{6}
\tablewidth{\columnwidth}
\tablecaption{[C/Fe] comparison with literature
}
\tablehead{   
  \colhead{ID} &
  \colhead{[C/Fe]$_{\text{This work}}$} &
  \colhead{[C/Fe]$_{\text{ref}}$} & 
  \colhead{$\Delta$[C/Fe]} &
  \colhead{Ref.}\\
  &
  \colhead{(dex)} &
  \colhead{(dex)} &
  \colhead{(dex)}
}
\startdata
S1020549 & $<0.25$ & $<0.20$ & $-$ & S15\\  
Scl11\_1\_4296 & $0.25\pm0.32$ & $0.34\pm0.34$ & $-0.09$ & S15\\
Scl07-50 & $<0.34$ & $-0.28\pm0.34$ & $-$ & S15\\
\noalign{\vskip 0.8mm} 
\hline
\noalign{\vskip 1.4mm} 
1008832 & $-1.14\pm0.27$ & $-0.88\pm0.10$ & $-0.26$ & K15\\
1007034 & $-1.01\pm0.37$ & $-1.11\pm0.10$ & $+0.10$ & K15\\
1007391 & $+0.55\pm0.38$ & $-0.05\pm0.13$ & $+0.60$ & K15\\
1009538 & $-0.78\pm0.61$ & $-0.80\pm0.11$ & $+0.02$ & K15\\
1010633 & $-0.84\pm0.31$ & $-0.84\pm0.10$ & $0.00$ & K15\\
1013035 & $< 0.00$ & $<-1.24$ & $-$ & K15\\
1013808 & $< 0.22$ & $-1.05\pm0.27$ & $-$ & K15\\
1016486 & $-0.26\pm0.36$ & $-0.65\pm0.12$ & $+0.39$ & K15\\
\noalign{\vskip 0.8mm} 
\hline
\noalign{\vskip 1.4mm} 
ET0381 & $-0.18\pm0.34$ & $-1.00\pm0.15^\dagger$ & $+0.82^\ddagger$ & J15\\
scl\_03\_059 & $-0.39\pm0.40$ & $-1.20\pm0.40\dagger$ & $+0.81^\ddagger$ & J15
\enddata
\tablecomments{S15, K15, and J15 refer to \citet{sjf+15}, \citet{kgz+15}, and \cite{jnm+15}, respectively.\\
$^\dagger$ \citet{jnm+15} present asymmetric uncertainties. These are the average of their asymmetric uncertainties.\\
$^\ddagger$ See Section~\ref{sec:jabcfe} for a discussion of the potential causes of these discrepancies.}
\label{tab:CFe}
\end{deluxetable}

\subsection{External Validation: Comparison to \citet{kgz+15} and \citet{sjf+15}}

Three stars in our sample have high-resolution [C/Fe] measurements in \citet{sjf+15}
with which we find agreement, as shown in Table~\ref{tab:CFe}.  Eight stars in our sample 
have medium-resolution [C/Fe] measurements in \citet{kgz+15}.
We find good agreement with their measurements, except for one star for which we measure 
a higher [C/Fe] by 0.6\,dex. If we adopt the stellar parameters provided by \citet{kgz+15}, then the
discrepancy reduces to 0.33\,dex. This resulting discrepancy appears to be reasonable given the
reported uncertainty in our [C/Fe] measurements of $\sim0.35\,$dex and the low S/N of the M2FS spectrum of the star.


\begin{figure}[!t]
\centering
\includegraphics[width = \columnwidth]{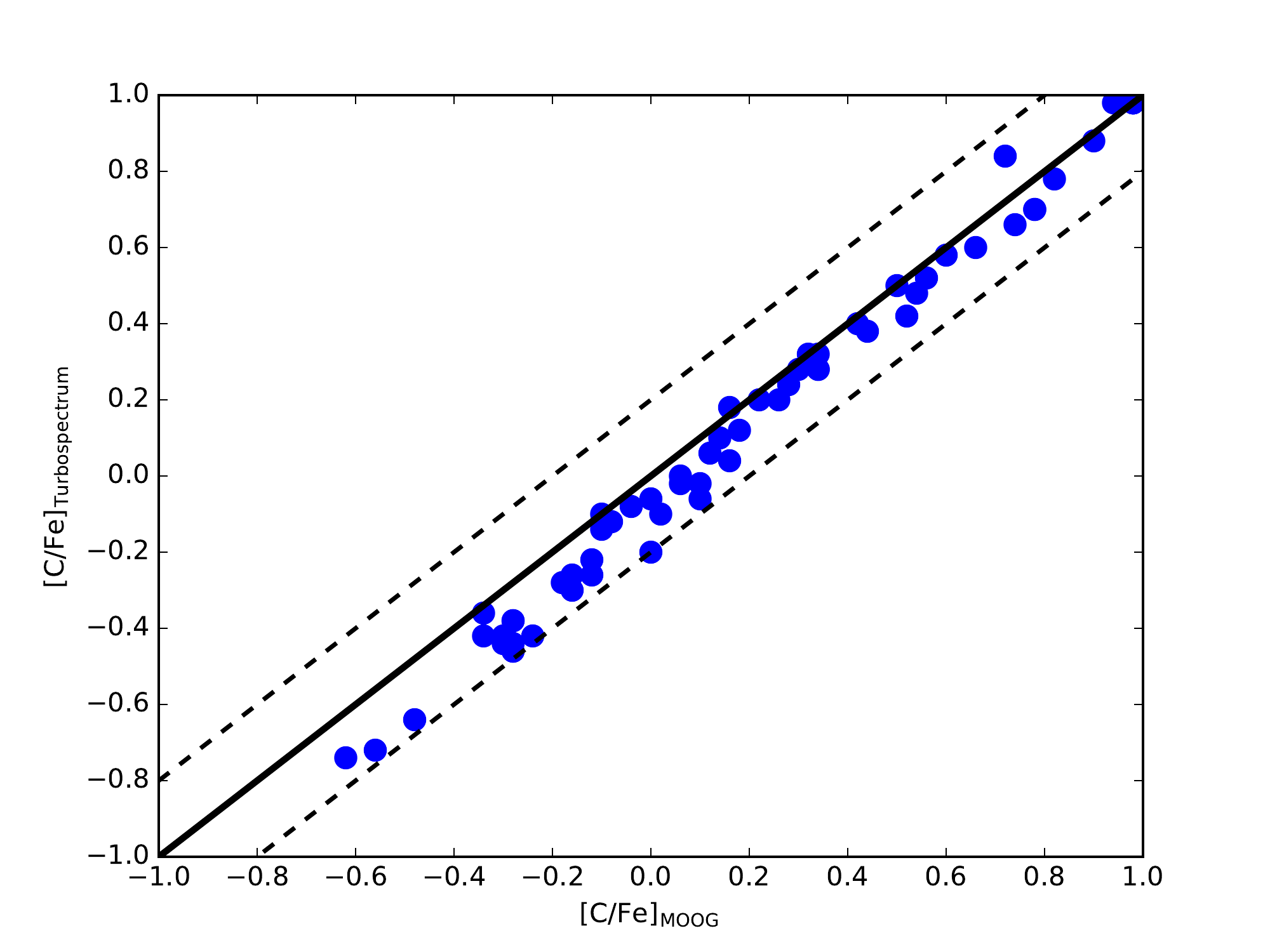}
\caption{[C/Fe] measured with Turbospectrum, the MARCS model
atmospheres, and the \citet{mpv+14} line list vs. [C/Fe] measured with MOOG and the same inputs. 
Dashed lines indicate $\pm0.2\,$dex offsets to guide the eye.}
\label{fig:SynthCompare}
\end{figure}


\subsection{External Validation: Comparison to \citet{jnm+15}}
\label{sec:jabcfe}
Two stars in our sample have high-resolution [C/Fe] measurements in \citet{jnm+15}.
We do not find agreement in [C/Fe] measurements, as our measurements are
at least $\sim0.8\,$dex higher (see Table~\ref{tab:CFe}). 
We note that \citet{jnm+15} adopted $\log\epsilon(\text{C})_{\odot} =  8.55$ \citep{ag+89, gs+98}, which is discrepant 
with the $\log\epsilon(\text{C})_{\odot} =  8.43$ assumed in MOOG \citep{ags+09}. 
This can account for 0.12\,dex of the total [C/Fe] offset between the measurements.

To explore whether the rest of this discrepancy could reasonably be explained by differences in the spectrum 
synthesis codes, model atmospheres, or line lists, we first attempted to reproduce the synthesis 
shown in \citet{jnm+15} for star ET0381. 
We were able to reproduce their synthesis using Turbospectrum, the MARCS model atmosphere, and the Masseron line list, but noticed a consistent offset of $\sim$0.5\,dex if we attempted to reproduce the synthesis with our adopted line list and MOOG.
This total observed discrepancy between our two approaches reasonably accounts for most of the observed
offset between [C/Fe] measurements, and about half of this observed $\sim0.5\,$dex discrepancy can be 
ascribed to differences in the line lists and adopted solar abundances.

To ensure that our CEMP detections were not susceptible to differences in synthesis codes, line lists, and model
atmospheres, we replicated our analysis for our CEMP stars using the two test grids discussed in Section~\ref{sec:synth}. 
We measured [C/Fe] for the subset of stars falling within the grid. As shown in Figure~\ref{fig:SynthCompare}, the discrepancies in [C/Fe] between the two synthesis codes are largely within 0.2\,dex, but grow larger for carbon-poor stars. 
Different model atmospheres and input line lists cause up to another $\sim0.1$ to $0.2$\,dex difference.
Referring to Figure~\ref{fig:SynthCompare}, we note that a star such as ET0381 with a measurement of [C/Fe]$\sim-0.20$ in MOOG tends to have an abundance lower by $\sim0.15$\,dex in Turbospectrum. If we apply additional offsets accounting for differences in line lists and adopted solar abundances, we recover the aforementioned offset of $\sim0.50$\,dex.
However, the classification of carbon-enhanced stars appears to be largely robust to different synthesis codes, model atmospheres, and input line lists.

\subsection{Confirmation of [C/Fe] with MagE spectra and further classification}
\label{sec:conf}

Motivated by the high number of CEMP stars
in the M2FS sample, we conducted follow-up observations of ten Sculptor stars with the MagE spectrograph as outlined in Section~\ref{sec:MagE}. This sample
included five strong CEMP candidates, and five stars that were not as carbon-enhanced but had similar stellar parameters to the five CEMP candidates.
We also observed one halo CEMP-r/s star, CS29497-034, as a comparison. 

\begin{figure}[!t]
\centering
\includegraphics[width = \columnwidth]{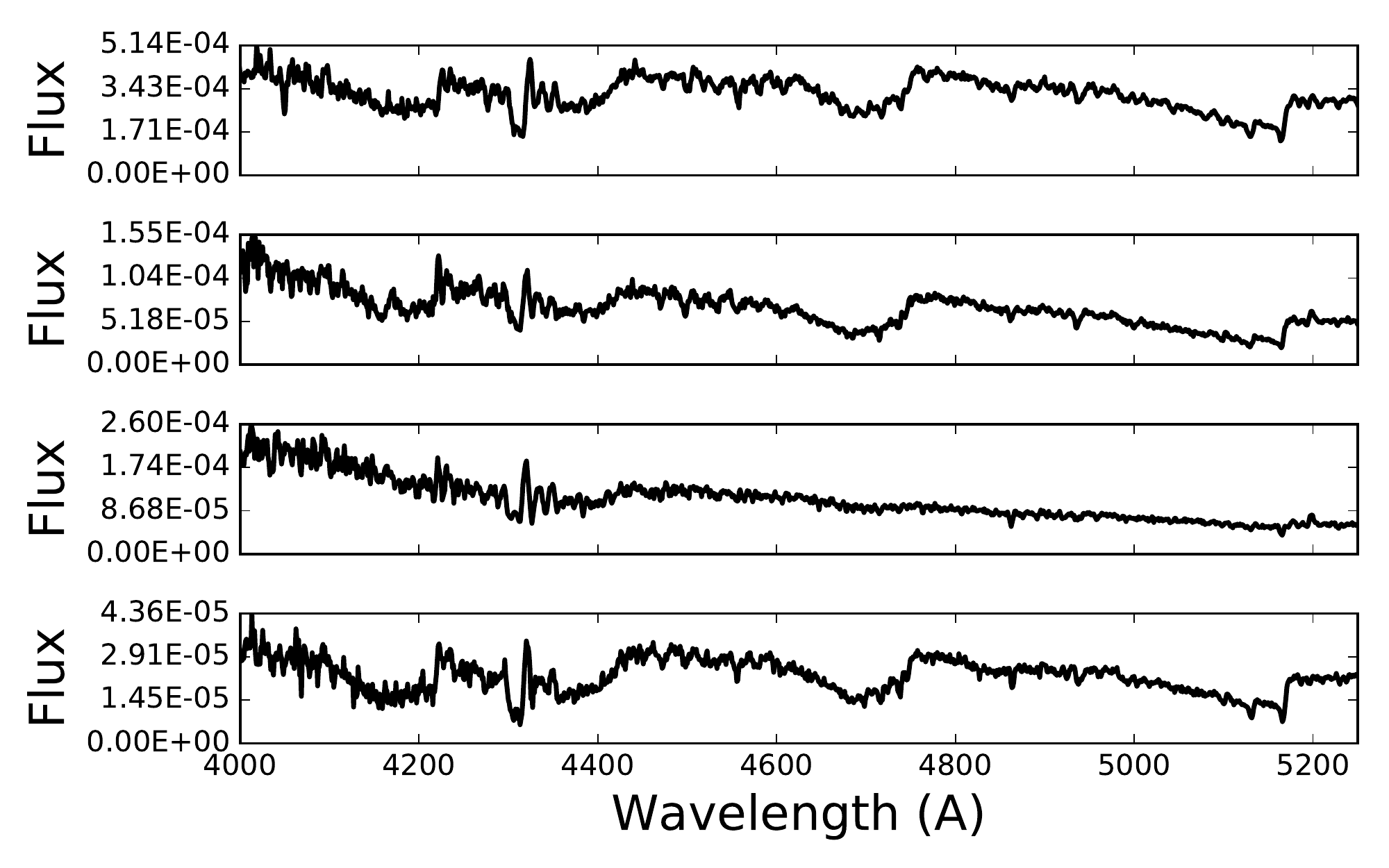}
\caption{M2FS spectra of 4 stars (from top: 10\_7\_486, 10\_8\_3963, 11\_1\_4121,
and 11\_1\_6440) that have saturated G-bands ($\sim4315$\,\AA). We measure their carbon abundance using the C2
band head at 5165\,\AA\,\,in their corresponding MagE spectra.}
\vspace{0.50cm}
\label{fig:cempspec}
\end{figure}

The purpose of these observations was to apply an independent check on our overall classification scheme,
and to potentially derive the barium abundance of the stars to further classify them.
Large Ba abundances in carbon-rich metal-poor stars are a strong indicator of the stars belonging to the CEMP-s and CEMP-r/s classes that 
are generally explained as being caused by accretion from a binary companion \citep{han+16b}. The more metal-rich analogs are 
the CH-strong and Ba-strong stars \citep{mw+90}. Any of these stars have to be excluded when computing a CEMP fraction, 
as their carbon enhancement does not reflect the abundance pattern in their birth environment. 
We indeed verified the carbon-rich nature of
the five stars in our sample, but found four of them to be more metal-rich stars of potentially 
either the CH-strong or Ba-strong class (see Section~\ref{sec:bafe}).
The other star was observed with the 1\arcsec\,\,slit, which does not provide sufficient resolution to measure barium features. 
M2FS spectra of a few strong carbon-enhanced stars are shown in Figure~\ref{fig:cempspec}.


\subsection{Identifying accreting binary carbon-rich stars in our M2FS sample}
\label{sec:crich}
\begin{figure}[!htbp]
\centering
\includegraphics[width =1.0\columnwidth]{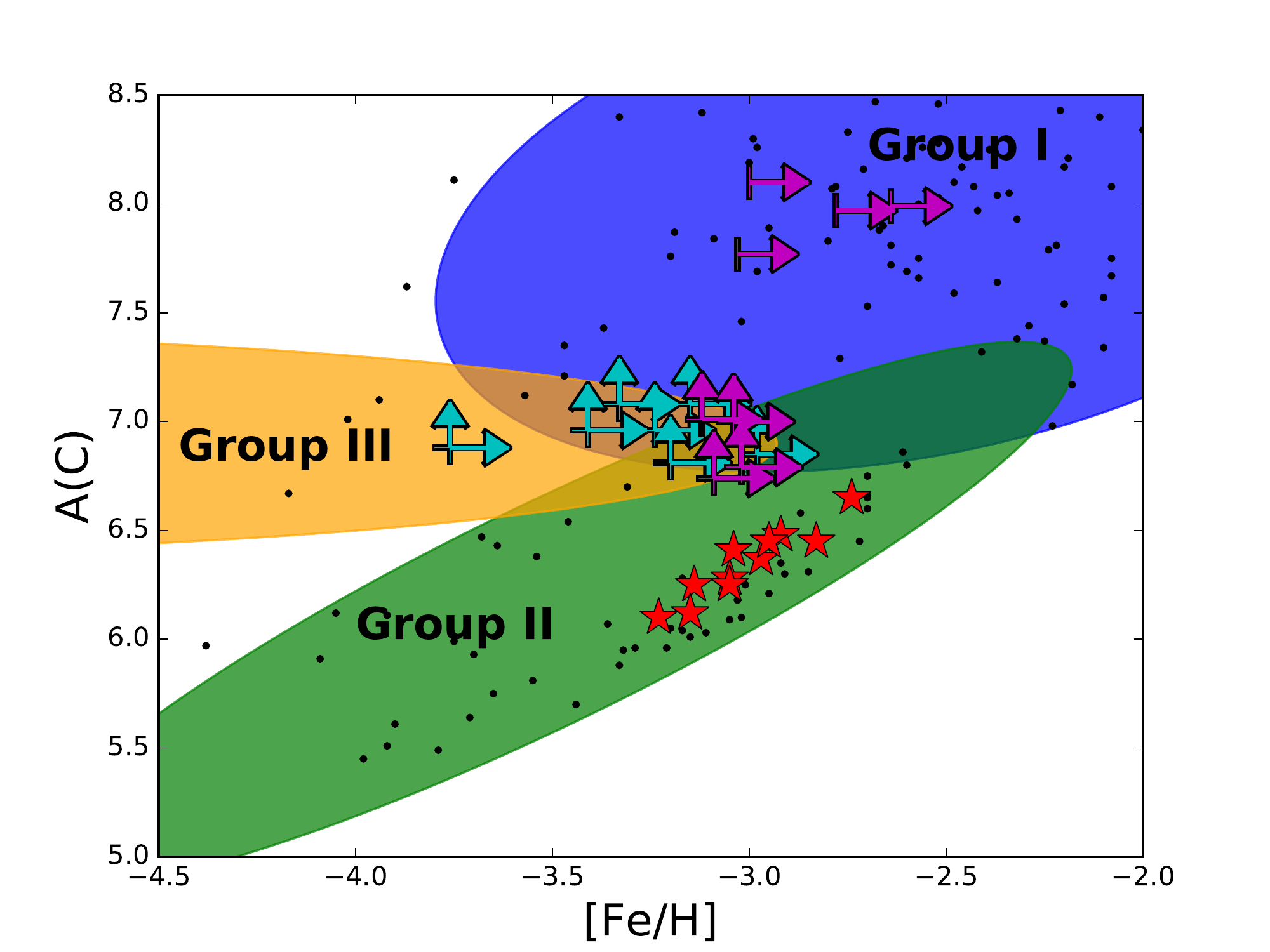}
\caption{Yoon et al. plot with the original sample of halo stars in black and our Sculptor CEMP candidates overlaid in red, cyan,
and magenta points. Groups I, II, and III are represented by blue, green, and orange ellipses, respectively.
Cyan points correspond to M2FS measurements of stars with saturated G-bands and lower limits on their carbon abundances and metallicities, 
magenta points correspond to M2FS measurements of stars with saturated G-bands but accompanying MagE carbon abundance measurements, and magenta points in Group I are MagE measurements of those stars with saturated G-bands. 
The majority of Group I stars are CEMP-s stars, and the majority of Group II and III stars are CEMP-no stars}
\label{fig:ybplot}
\end{figure}

It is necessary to exclude carbon-rich stars whose  source of enhancement is extrinsic (e.g. accretion from a binary companion) from  our calculation of the CEMP fraction. 
Generally, members of this class of carbon-rich binary stars can be identified by radial velocity monitoring or by detecting a combined enhancement in s-process elements (e.g. Ba) together with carbon that would have been produced in a companion asymptotic giant branch star.
But recent work by \citet{ybp+16} suggests that stars with sufficiently high absolute carbon abundance (A(C))
can already be identified as CEMP-s stars just based on the [Fe/H] and A(C) measurements, as shown in Figure~\ref{fig:ybplot}.

We can readily apply the Yoon et al. criterion to both our M2FS and MagE samples. However, for the four most carbon-enhanced stars in our MagE sample there is a discrepancy in our carbon abundance measurements. The A(C) values derived from the MagE data suggests these stars to be  clearly s-process rich stars, while the M2FS A(C) measurements place them on the boundary according to the Yoon et al. criterion. 

The higher resolution of the MagE spectra better resolves the G-band and the C$_2$ band head and suggests that these four stars are more carbon-enhanced than inferred from the lower resolution M2FS spectra.
In addition, renewed inspection of the Ca II K line reveals the same trend; these four stars are actually more metal-rich than the KP index measurement from the M2FS data had indicated.
Overall, these revisions strongly suggest that the four stars could be either CEMP-s stars (if they indeed have [Fe/H] 
$\lesssim -1.5$), or belong to the class of even more metal-rich CH-strong or Ba-strong stars.


\begin{figure*}[!t]
\centering
\includegraphics[width = \textwidth]{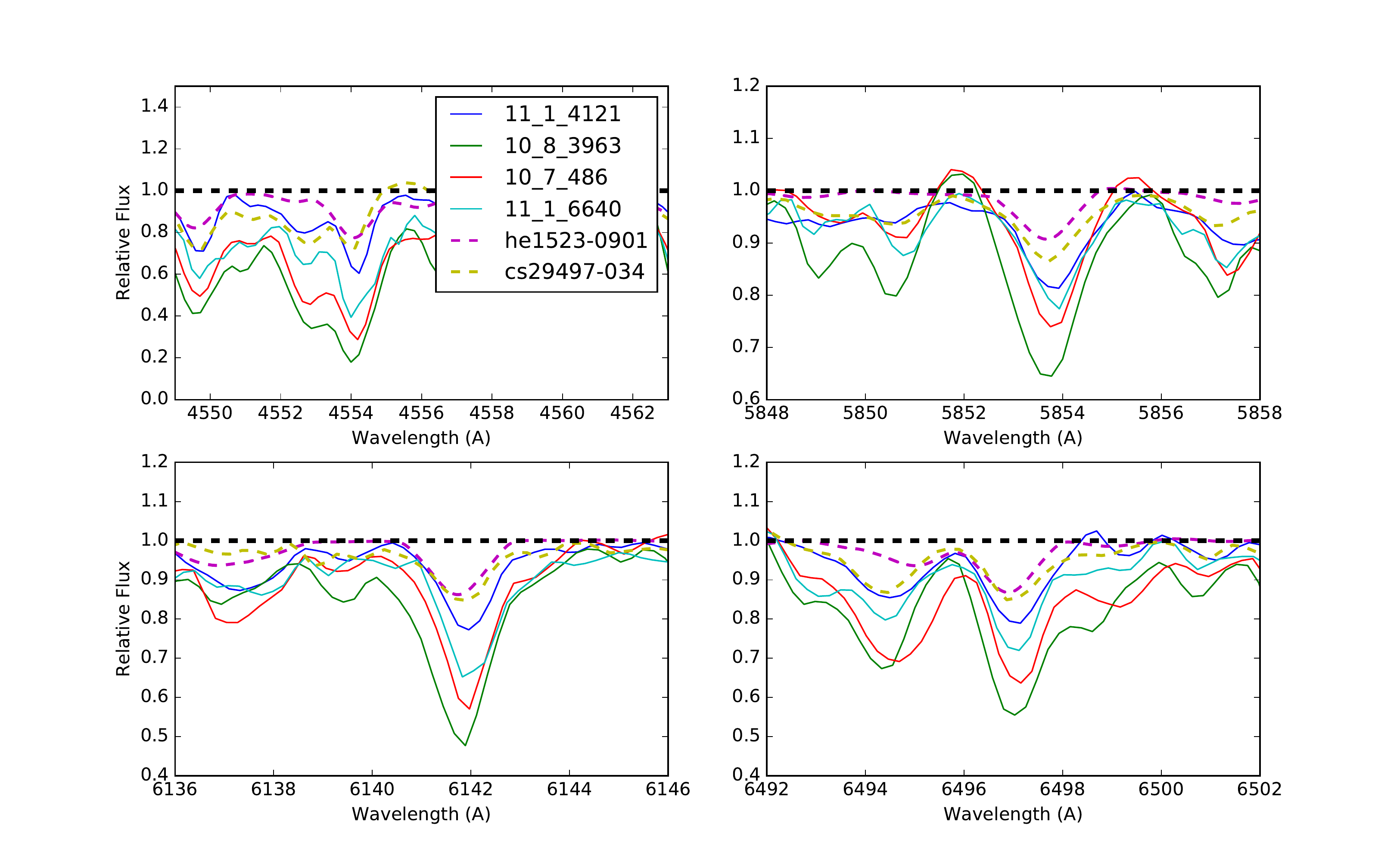}
\caption{Plots of barium lines at 4554\,\AA, 5853\,\AA, 6141\,\AA, and 6496\,\AA\,\,in MagE $R\sim6000$ spectra for 4 Sculptor CEMP 
stars (solid lines). The MagE ($R\sim6000$) spectrum of CS29497-034 ([Ba/Fe] = $2.23$ from \citealt{abc+07}), a halo CEMP-r/s star, 
and a high-resolution MIKE spectrum of HE1523-0901 ([Ba/Fe] $\sim 1.1$ from \citealt{fcn+07}), an
r-process enhanced star, smoothed to $R\sim6000$ are over plotted for comparison.}
\vspace{0.50cm}
\label{fig:balines}
\end{figure*}


Regarding the carbon abundance discrepancy, we note that when high carbon abundances lead to strong spectral absorption features (especially in cool stars), there is no region in the vicinity of the G-band ($4250$\,\AA\,\,to $4350$\,\AA) 
to place the true continuum value in M2FS spectra. Thus, even accounting for this effect can still easily lead to systematically underestimating the continuum, and thus the carbon abundance. These four stars all had [C/Fe]$_{\text{M2FS}} \gtrsim 1$. We thus speculate that the G-band in M2FS spectra begins to saturate around [C/Fe]$_{\text{M2FS}} \sim 1$.

We note that the G-band in the higher-resolution MagE spectra also begins to saturate for those four stars. This is illustrated by our inability to use the G-band to recover the literature [C/Fe] measurement of CS29497-034, a star with similar G-band depth in the MagE spectra as our Sculptor members with high [C/Fe]. Motivated by the near-saturation of the G-band for these stars, we instead determined the carbon abundances of CS29497-034 using the C$_2$ band head at 5165\,\AA. We used a line list compiled from \citet{slc+09, sck+16} and \citet{mpv+14} and the MOOG synthesis code. We measure [C/Fe] = 2.6$\pm0.1$ for CS29497-034, consistent with the literature value of [C/Fe] = 2.72 \citep{wbc+07}. We thus use the C$_2$ band head to measure carbon abundances for the stars observed with MagE that have a near saturated G-band.

We find 11 stars with [C/Fe]$_{\text{M2FS}} > 1.0$ and showing the presence of a C$_2$ band head and a very strong G-band, which we suspect to have underestimated carbon abundances. If the A(C) value of these stars were revised upwards by $\sim$1\,dex (following the results for CS29497-034 and the four stars also observed with MagE), they would clearly be members of the class of s-process rich stars, based on the Yoon et al. plot (see Figure~\ref{fig:ybplot}). 
We thus consider these stars as s-process rich candidates, and list our derived carbon abundances strictly as lower limits in Table~\ref{tab:M2FS} and Figure~\ref{fig:CEMPfracs}.
Table~\ref{tab:MagE} has a final list of the iron and carbon abundances computed 
for the subset of all 31 MagE spectra with $B-V < 1.2$. 
Given the ambiguity in the metallicities of the carbon-rich stars observed with MagE, we cautiously only list A(C) measurements for those stars.

\section{Chemical signatures of the metal-poor stellar population of Sculptor}
\label{sec:additional}

\subsection{[Ba/Fe] estimates from MagE spectra \& exclusion from CEMP-no classification}
\label{sec:bafe}

In our follow-up MagE observations (Section~\ref{sec:MagE}) of ten stars, we observed four of the five very carbon-enhanced candidates with the 0\farcs7\,slit 
to obtain sufficient resolution ($R\sim6000$) to also resolve barium lines 
at 4554\,\AA, 4934\,\AA, 5853\,\AA, and 6141\,\AA. 
We used a line list from \citet{slc+09, sck+16} and the MOOG 
synthesis code to synthesize these lines and constrain [Ba/Fe].

At $R\sim6000$, these four lines can be blended, e.g., with praseodymium at 5853\,\AA, when neutron-capture element abundances are high as in s-process-rich stars. 
We are able to reproduce the literature [Ba/Fe] = 2.2 measurement of CS29497-034 when considering the depth of the centroid of the line and neglecting fitting the entire line profile. 
This suggests that the blending features do not significantly affect the centroid of the barium lines. 
In Figure~\ref{fig:balines}, the barium lines of the stars are over-plotted with the resolution-degraded MIKE spectrum of the halo r-process star HE~1523$-$0901 \citep{fcn+07}, which has similar stellar parameters to the four Sculptor stars. 
The barium features of the Sculptor stars are stronger than those in the reference stars CS29497-034 ([Ba/Fe] = 2.2) and HE~1523$-$0901 ([Ba/Fe] = 1.1), suggesting that they are s-process enhanced stars with [Ba/Fe] $>$ 1.0. 
The centroid measurements for these stars yield high [Ba/H] values of 0.36, 0.8, $-0.53$, and $-0.18$. Taking our KP-based Fe measurements at face value, these abundances translate to [Ba/Fe] = 3.00, 3.80, 2.50, and 2.60. However, these stars show strong CH features in the vicinity of the Ca II K line in their spectra. This prevents an accurate [Fe/H] measurement (see Section~\ref{sec:resKP}). Even if the [Fe/H] values of these stars were underestimated by up to 1.5\,dex, these stars would still be considered s-process rich stars due to their high barium abundance. In addition, just based on the A(C) criteria described in \citet{ybp+16}, and as shown in Figure~\ref{fig:ybplot},
these stars could independently be classified as s-process rich stars.


\begin{figure*}[!t]
\centering
\includegraphics[width = \textwidth]{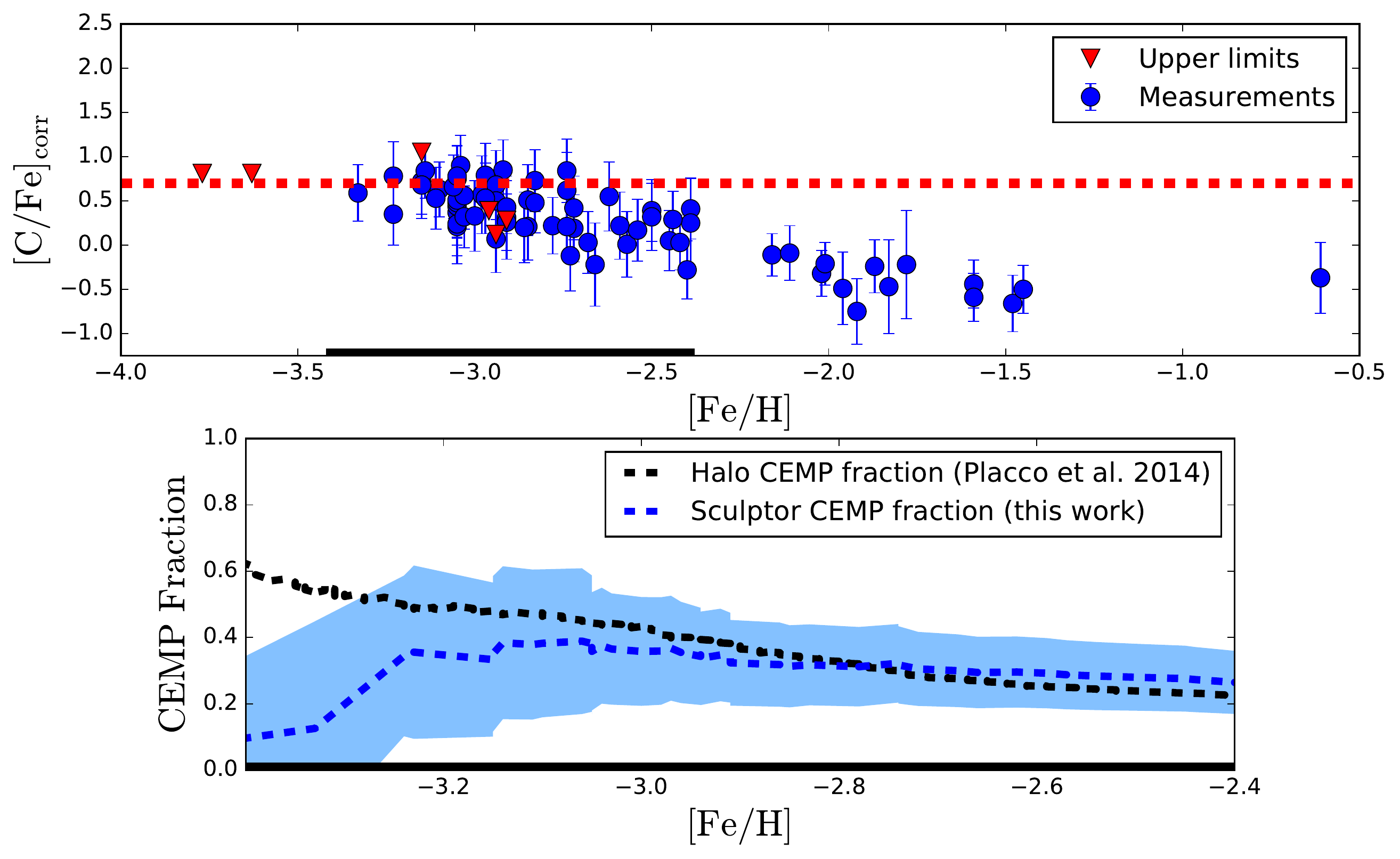}
\caption{Top: [C/Fe] as a function of [Fe/H] for RGB stars in our M2FS Sculptor sample. CH strong, Ba strong, and CEMP-s candidates are not displayed in the upper panel of the plot. The displayed [C/Fe] measurements
have been corrected for the evolutionary state of each star following \citet{pfb+14}. The dashed 
red line marks the cutoff for a star to be considered a CEMP star ([C/Fe] $>$ 0.7). Red downward-facing triangles are upper limits on [C/Fe]
from non-detections of the G-band.
Bottom: Measured cumulative CEMP fraction as a function of [Fe/H] for our Sculptor sample (blue) and the Milky Way
halo from \citet{pfb+14} (black). The shaded blue region corresponds to the 95\% confidence interval of our measured CEMP
fraction.}
\vspace{0.50cm}
\label{fig:CEMPfracs}
\end{figure*}


\subsection{Sample bias assessment}
\label{sec:resKP}

Our M2FS sample is composed of the most metal-poor members of Sculptor as selected from measurements of the Ca II K line in lower-resolution IMACS spectra. Our initial metallicity cut based on the IMACS data attempted to include all stars with [Fe/H] $< -2.9$.
The majority of stars are cool red giants. 
There is a potential for CEMP stars to be preferentially included or excluded from the M2FS sample if their metallicity measurements are systematically biased because of strong C absorption. 
At face value, we expect CH absorption features to depress the continuum blueward of the Ca II K line in the lower-resolution IMACS spectra, causing a lower measurement of the equivalent width of the Ca II K line and thus a faulty selection. 
This would mean that carbon-rich stars may be preferentially selected into our M2FS sample because they may appear to be extremely metal poor stars. 

\begin{figure}[!htbp]
\centering
\includegraphics[width =1.0\columnwidth]{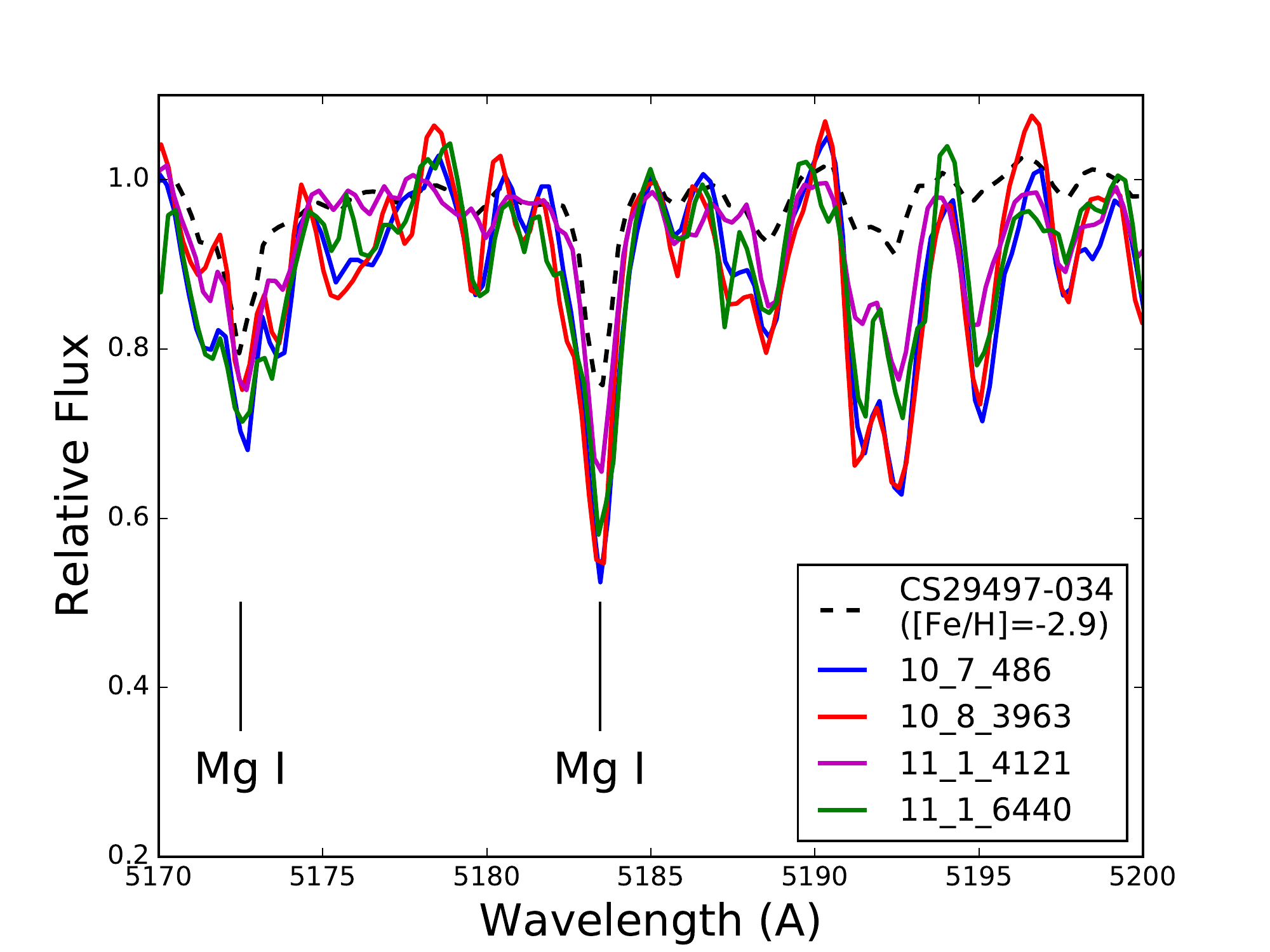}
\caption{Plot of the Mg region of the MagE spectra of CS29497-034 ([Fe/H] = $-2.9$) and four other more metal-rich Sculptor members. 
These stars were classified as [Fe/H]$\sim -3.0$ from measurements of the Ca II K line. 
It appears that the strong carbon-enhancement of these Sculptor members biased the Ca II K metallicities in lower-resolution spectra (see Section~\ref{sec:resKP}).}
\label{fig:mgregion}
\end{figure}

Stars whose carbon-enhancement is driven by accretion across a binary system, such as CEMP-s, Ba-strong, and CH-strong stars, have the highest A(C) values and would thus be the most likely to be preferentially selected into our sample.
Indeed, we find 4 more metal-rich CEMP-s, Ba-strong, or CH-strong stars in our M2FS sample based on
follow-up observations with MagE (see Section~\ref{sec:crich}).  
All of these stars were initially found to have [Fe/H]$\sim -3.0$ based on measurements of the strength of the Ca II K line.
But these stars must actually be much more metal-rich as a simple comparison of the magnesium triplet region ($\sim5175$\AA) of these stars to that of the halo CEMP-r/s star CS29497-034 ([Fe/H] = $-2.9$) shows (see Figure~\ref{fig:mgregion}). 
Given this comparison, we also chose to investigate the magnesium triplet region of stars without extreme A(C) values to determine whether their metallicity measurements were biased.

For each star in Table~\ref{tab:MagE}, we derived a Mg abundance from the 5172.7\,\AA\,\,and 5183.6\,\AA\,\,lines if the S/N was sufficiently high.
Then, we compared the derived [Mg/Fe] ratio of these stars to the expected [Mg/Fe] ratio for dwarf galaxy stars in their metallicity regime.
We would expect to see systematically higher [Mg/Fe] values if the Ca II K based metallicities were biased lower, such as in the case of stars with high A(C) values.

We consider two examples: stars 10\_7\_442 and 10\_8\_1226 have carbon abundances close to the CEMP threshold and Mg line equivalent widths in the linear regime of the curve of growth (reduced equivalent widths $\lesssim -$4.45). 
For these two stars, we measure [Mg/Fe] values of 0.23 and 0.17, respectively. 
These [Mg/Fe] ratios are roughly at the lower end of the regime of what is expected for dwarf galaxy stars at these metallicities. 
This suggests that we are not strongly underestimating our [Fe/H] measurements for stars that are near the CEMP threshold.

If we include stars from Table~\ref{tab:MagE} with Mg line equivalent width measurements in the non-linear regime of the curve of growth at face value and adopt the M2FS metallicities and carbon abundances when available, the average [Mg/Fe] of stars with [C/Fe] $>$ 0.50 is 0.43. 
This [Mg/Fe] ratio is also in the regime of expected values. 
As mentioned, if the metallicities were substantially underestimated, we would expect to get much larger [Mg/Fe] values. 
For comparison, all the CEMP-s candidates have [Mg/Fe] $\gtrsim$ 1.0 if we take the KP-based [Fe/H] measurements at face value.
While these Mg abundance estimates may have large uncertainties (up to $\sim0.4$\,dex, as is expected for data of this quality), they suggest we are not strongly biased in our metallicity estimates for stars without copious carbon-enhancement.

We also compared our observed MagE spectra to MIKE spectra of CS22892-52 ([Fe/H] = $-3.16$; $T_{\text{eff}}$ = 4690\,K) and HD122563 ([Fe/H] = $-2.93$; $T_{\text{eff}}$ = 4500\,K) that had been degraded to match the resolution of the MagE data.
Measurements of these standard stars are from \citet{rpt+14}.
We find that the strengths of the Mg b lines observed with MagE appear to be roughly consistent with what is expected from our Ca II K derived metallicities.  

Thus, only stars with very strong carbon enhancement are incorrectl selected into our M2FS sample. 
These stars are overwhelmingly likely to have their carbon abundance elevated by accretion from a binary companion (see Figure~\ref{fig:ybplot}), and should already be excluded in a calculation of the CEMP fraction.
This confirms that our selection is not biased in favor of CEMP-no stars.

Below a fiducial metallicity of [Fe/H] $\sim -3.0$ and after excluding CEMP-s, Ba-strong, and CH-strong stars, we can reasonably assume that there is not a strong bias toward high carbon enhancement in our EMP sample in Sculptor.

\begin{figure}[!t]
\centering
\includegraphics[width = \columnwidth]{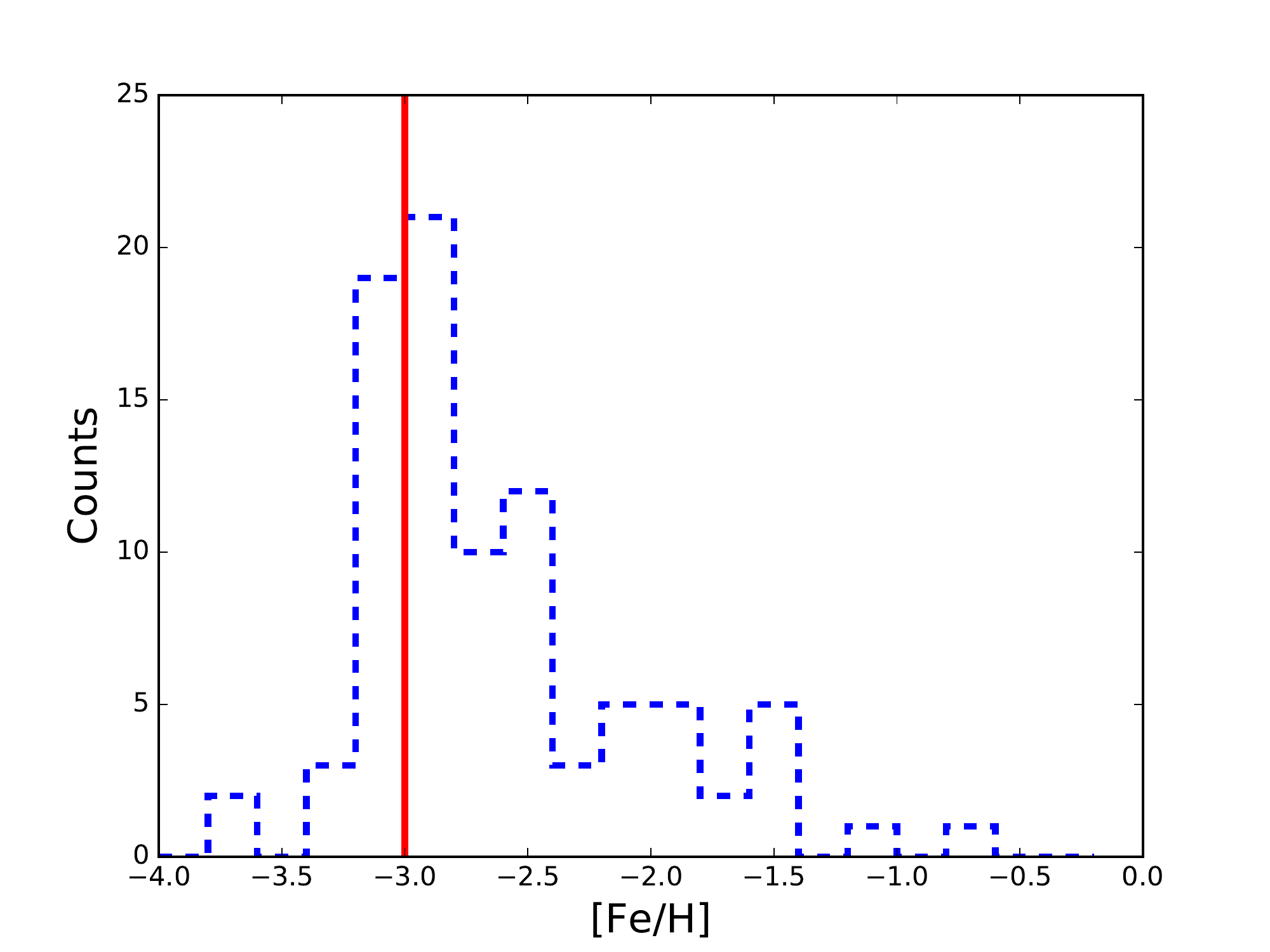}
\caption{Histogram of the metallicities measured for 89 stars. Star with lower limits on metallicities are not included. The vertical red line indicates the cutoff for
extremely metal-poor stars ([Fe/H]$ < -3.0$). After excluding lower limits on [Fe/H], we detect 24 extremely metal-poor star candidates.}
\label{fig:fehhist}
\end{figure}


\subsection{Measurement of the CEMP fraction in Sculptor}
\label{sec:measurement}

In a measurement of the CEMP fraction, we must exclude stars whose carbon enhancement is extrinsic (e.g. driven by accretion from a binary companion). 
We identify such stars in our M2FS sample by applying the Yoon et al. criterion (see Figure~\ref{fig:ybplot}), as discussed in Section~\ref{sec:crich} and Section~\ref{sec:bafe}. 
We then excluded 90\% of those stars, which is the probability of correct classification according to Yoon et al., from our calculation of the CEMP fraction. 

We note that there is a group of stars that sits blueward of the Sculptor RGB by up to $\sim$0.25\,mag (see Figure~\ref{fig:CMD}). 
Despite detailed investigation, the evolutionary status and hence the nature of these stars remains somewhat ambiguous. 
While they are generally bluer than would be expected for Sculptor RGB stars, they do tend to have velocities similar to Sculptor.
Due to this uncertainty, we thus cautiously exclude these stars from our calculation of the CEMP fraction and we list them separately in Table~\ref{tab:M2FS}. 
Since they comprise only a small portion of the sample, the CEMP fraction is largely unchanged by their exclusion.  

We determined the CEMP fraction by accounting for the probability that any individual star in our sample is 
carbon enhanced ([C/Fe] $>$ 0.7). We assigned a probability that each star is carbon enhanced based
on its [C/Fe] measurement and assuming that the uncertainty on [C/Fe] is normally distributed. 
Finally, we computed a cumulative CEMP fraction for each metallicity range 
by finding the expected number of CEMP stars in that subset based on the
probabilities of each member being carbon enhanced. 
We then divided the expected number
of CEMP stars by the total number of stars in the subset. 
This approach enables us to accurately constrain the overall population of such stars even though we are not able to identify individual CEMP stars with high (p$>$0.95) confidence.

To derive an uncertainty on this CEMP fraction, we modeled the CEMP classification as a random walk where 
$p_i$ is the probability of a given star being a CEMP star. This formulation yields an uncertainty on the CEMP fraction of 
$\Sigma_i\sqrt{p_i\times(1-p_i)/N}$. This uncertainty matches the uncertainty derived from Monte Carlo simulations
of the CEMP fraction. We can then measure the observed cumulative CEMP fraction, and an uncertainty on the fraction,
as a function of [Fe/H]. 
While abundance uncertainties in [C/Fe] for our sample are almost certainly non-Gaussian 
as the distributions of $T_{\text{eff}}$ and $\log g$ residuals with respect to \citet{kcg+13} 
are asymmetric, this method allows us to place a rough uncertainty on the observed CEMP fraction.
The results of this analysis are shown in Figure~\ref{fig:CEMPfracs}.
To test the impact of non-gaussianity, we compiled the individual [C/Fe] measurements that were used to calculate the uncertainty on the final carbon abundance of each star (see Section~\ref{sec:gridfit}).
We then calculated the fraction of those [C/Fe] measurements that were greater than 0.7\,dex for stars with [Fe/H] $< -3.0$.
We find that the fraction of those [C/Fe] measurements ($\sim40$\%) is in agreement with our final reported CEMP fraction for EMP stars.

We measure a CEMP fraction of $0.36\pm0.08$ for stars below [Fe/H] = $-$3.0 in Sculptor.
If we instead take the simpler approach of dividing the number of CEMP stars by the total
number of stars after excluding 10 of the 11 probable CEMP-s stars, we derive a CEMP fraction of $0.24$ (6/25) for stars below [Fe/H] = $-3.0$. 
The latter fraction is likely lower because our carbon abundances have large uncertainties ($\sim0.35$\,dex) and a number of stars lie right below the cutoff of the CEMP classification. 
Exactly this fact has been taken into account by the method described in the previous paragraph, so we adopt the former measurement. 

\begin{deluxetable*}{lllllrrrrrr}
\tablewidth{1.0\textwidth}
\tablecolumns{7}
\tablecaption{Stellar parameters and abundances from MagE spectra}
\tablehead{   
  \colhead{Names} &
  \colhead{Slit} &
  \colhead{Log (g)} &
  \colhead{Teff} &
  \colhead{[Fe/H]$_{\text{KP}}$} &
  \colhead{A(C)} &
  \colhead{[C/Fe]} &
  \colhead{[C/Fe]$_{\text{corr}}$}&
  \colhead{[C/Fe]$_{\text{final}}$}&
  \colhead{[Ba/H]}  \\
   	\colhead{} &
  \colhead{(arcsec)} &
  \colhead{(dex)} &
  \colhead{(K)} &
  \colhead{(dex)} &
  \colhead{(dex)} &
    \colhead{(dex)} &
  \colhead{(dex)} &
  \colhead{(dex)} &
  \colhead{(dex)} 
}
CS29497$-$034$^\dagger$ & 0.7 & 1.50 & 4900 & $-2.90\pm0.27$ & $8.25\pm0.29$ & $2.60\pm0.10$  & 0.09 & $2.69\pm0.10$ & $-0.70$:$^\ddagger$  \\
10\_8\_3963 & 0.7 & 1.08 & 4513 & $>-3.00$ & $8.10\pm0.15$ & ... & ... & ...& 0.80:$^\ddagger$ \\
10\_7\_486 & 0.7 & 1.05 & 4523 & $>-2.64$ & $7.96\pm0.15$ & ... & ... & ... & 0.36:$^\ddagger$ \\
11\_1\_6440 & 0.7 & 1.29 & 4605 & $>-2.78$ & $7.82\pm0.15$ & ... & ... & ... & $-0.18$:$^\ddagger$  \\
11\_1\_4121 & 0.7 & 1.24 & 4579 & $>-3.03$ & $7.52\pm0.10$ & ... & ... & ... & $-0.53$:$^\ddagger$ \\
11\_1\_4422 & 1.0 & 1.75 & 4810 & $-2.85\pm0.23$ & $6.80\pm0.34$ & $1.10\pm0.25$ & 0.16 & $1.26\pm0.25$ &...  \\
6\_5\_1598 & 1.0 & 1.08 & 4516 & $-2.83\pm0.16$ & $6.02\pm0.26$ & $0.30\pm0.20$ & 0.65 & $0.95\pm0.20$ & ... \\
11\_2\_661 & 1.0 & 1.16 & 4550 & $-2.93\pm0.17$ & $5.67\pm0.23$ & $0.05\pm0.15$ & 0.67 & $0.72\pm0.15$ & ...   \\
10\_8\_1566 & 1.0 & 1.53 & 4659 & $-2.11\pm0.34$ & $5.84\pm0.40$ & $-0.60\pm0.20$ & 0.47 & $-0.13\pm0.20$ & ... \\
7\_4\_2408 & 1.0 & 1.06 & 4524 & $-2.64\pm0.16$ & $5.51\pm0.26$ &  $-0.40\pm0.20$ & 0.72 & $0.32\pm0.20$ & ...  \\
11\_1\_4673 & 1.0 & 1.21 & 4570 & $-2.94\pm0.18$ & $5.31\pm0.27$ & $-0.30\pm0.20$ & 0.65 & $0.35\pm0.20$ & ... \\
\noalign{\vskip 1mm} 
\hline
\noalign{\vskip 1mm} 
\hline
\noalign{\vskip 1mm} 
10\_8\_3804   & 1.0 & 1.62  &    4752 &  $>-2.78$ & $8.24\pm0.22$ & ... & ... & ... & ... \\
11\_1\_3334$^\dagger$ & 1.0 & 1.62 & 4721 & ... & $7.88\pm0.15$ & ... & ... & ... & ...  \\
6\_5\_505$^\dagger$ & 1.0 & 1.57 & 4706 & ... & $7.52\pm0.15$ & ... & ... & ... & ... \\
11\_2\_556  & 1.0  & 2.04 &     4939  & $>-3.27$  & $7.48\pm0.20$ & ... & ... & ... & ... \\ 
7\_4\_3280     & 0.7 & 3.59 & 5518   &  $-2.41\pm0.25$  & $<6.84$ & $<0.70$ & 0.00 & $<0.70$ & ... \\ 
10\_8\_2714  & 1.0 & 3.02 &    5328  &  $-2.96\pm0.38$   & $<6.59$ &  $<1.00$ & 0.01 & $<1.01$ & ... \\ 
10\_8\_3810   & 1.0  & 2.69 &    5199 & $-3.10\pm0.33$ & $<6.15$  &  $<0.70$ &  0.01 & $<0.71$ & ...\\ 
6\_5\_1035 & 0.7 &   1.27  &   4589 &  $-2.86\pm0.20$   & $5.69\pm0.28$ & $0.00\pm0.20$ &  0.61 & $0.61\pm0.20$  & ...\\
10\_8\_1226  & 1.0 &  1.47 & 4685   &  $-3.05\pm0.21$  & $5.68\pm0.33$ & $0.18\pm0.25$   & 0.44 & $0.62\pm0.25$ & ... \\ 
10\_7\_442   & 1.0  & 1.61 &    4752  & $-3.33\pm0.22$ & $5.67\pm0.30$ & $0.45\pm0.20$   & 0.29 & $0.74\pm0.20$ & ... \\
7\_4\_1992   & 1.0 & 1.66  &     4769  & $-3.14\pm0.22$  & $5.60\pm0.33$ &$0.19\pm0.25$   &  0.23 & $0.42\pm0.25$ & ... \\ 
11\_1\_4296 & 1.0  & 1.52 & 4720  & $-3.99\pm0.22$ & $<5.56$  & $<1.00$ & 0.36 & $<1.36$ & ...\\
11\_1\_6015  & 1.0 & 1.87  &   4824 & $-2.42\pm0.30$  & $5.53\pm0.36$ & $-0.60\pm0.20$  & 0.12 & $-0.48\pm0.20$ & ... \\ 
10\_7\_790  & 0.7  & 1.23 &     4574  & $-3.03\pm0.17$  & $5.47\pm0.34$ & $-0.05\pm0.30$  & 0.63 & $0.58\pm0.30$ & ... \\ 
6\_6\_402  & 1.0 &   1.68 &   4802 & $-3.91\pm0.25$  & $<5.44$ & $<0.80$  & 0.17 & $<0.97$  & ... \\ 
10\_7\_923    & 1.0  & 1.39  & 4666  & $-3.87\pm0.20$  & $<4.88$ & $<0.20$   & 0.49 & $<0.69$  & ...
\enddata
\tablecomments{Stellar parameters and [Fe/H] for CS29497$-$034 are from \citet{abc+07}. Stars in the top portion were observed
as a follow-up to M2FS observations to confirm [C/Fe] measurements, and stars in the bottom portion were observed immediately after the initial IMACS observations as EMP candidates.\\
$^\dagger$The S/N over the Ca II K feature was too low to estimate a [Fe/H] from the KP index. The M2FS [Fe/H] was assumed when
calculating [C/Fe] (See Table~\ref{tab:M2FS})\\
$^\ddagger$The colon (:) indicates large and uncertain error bars}.
\label{tab:MagE}
\end{deluxetable*}

\section{Discussion and Conclusion}
\label{sec:discussion}

The overall aim of this study has been to establish the early chemical evolution of Sculptor by studying
a sample of metal-poor stars in this galaxy. In particular, we obtained metallicity ([Fe/H]) and 
carbon abundance ([C/Fe]) measurements for 100 metal-poor stars in Sculptor using medium-resolution 
M2FS spectroscopy. We identify 21 carbon-enhanced metal-poor star candidates 
(CEMP; $\mbox{[C/Fe]} > 0.7$, \feh\, $<$ $-1.0$), and 24 extremely metal-poor 
candidates (EMP; \feh\, $<$ $-$3.0). The MDF of our sample is shown in Figure~\ref{fig:fehhist}.
Note that this sample is selected to have [Fe/H] $\sim -3$ and is not representative of the galaxy as a whole.

We also observed 31 stars with the MagE spectrograph of which 26 had $B-V < 1.2$. 
For ten, their carbon-enhanced nature was confirmed, enabling further insight into the origin of their carbon enhancement.

From these observations, we determine that many of our carbon-rich stars may be CEMP-s, CH-strong, or Ba-strong stars (see Sections~\ref{sec:crich} and ~\ref{sec:bafe}) but such stars should be excluded in an estimate of the CEMP fraction. Excluding 90\% of these stars, which is an approximation of their recovery rate, 
suggests a true CEMP fraction of 36\% (see Section~\ref{sec:measurement})
for EMP stars in Sculptor (see Figure~\ref{fig:CEMPfracs}).

Prior to this study, only four CEMP stars had been identified in Sculptor \citep{sts+15,lbp+16,sdy+16}. 
Of those, only one was a CEMP-no star, resulting in an apparent disagreement between the CEMP fraction
of Sculptor and the CEMP fraction of the Milky Way halo ($\sim42\%$). This discrepancy, if true, would have hinted at  
a divergence of the earliest phases of chemical evolution, as reflected in the most metal-poor stars
in the halo and in Sculptor. However, our CEMP fraction of $\sim36\%$ for EMP stars in Sculptor is in agreement with the CEMP
fraction of $\sim42\%$ for EMP stars in the Milky Way halo, posing
no such challenges.
This measurement is also consistent with theoretical predictions for the early evolution of Sculptor \citep{sst+15}.

In fact, our results show that Sculptor may have a similar cumulative CEMP fraction as the halo for 
stars with [Fe/H] $< -$3.0 (see Figure~\ref{fig:CEMPfracs}), using the compilation of metal-poor 
halo stars from \citet{pfb+14} for comparison. At face value, Figure~\ref{fig:CEMPfracs} suggests that Sculptor and the halo have the same CEMP fraction at all metallicities below [Fe/H] = $-2.5$. However, the large number of stars in our sample with [Fe/H] $\sim -2.8$ biases the measurement of the cumulative CEMP fraction towards the value at metallicities lower than that number.
Contrary to previous work, this suggests that a high CEMP star fraction may be a defining 
characteristic of the low metallicity Sculptor population after all, and also
suggests that in Sculptor, early chemical evolution was driven by high [C/Fe] 
producing objects such as fallback supernovae with large [C/Fe] yields and/or massive rotating stars with large 
CNO yields \citep{lcb+03, un+03, iut+05, h+07, mhe+10, hw+10, jab+10, tin+14}.

This result indicates that the earliest stars in Sculptor and, perhaps, more generally in all classical dSphs, may have undergone similar processes of early chemical enrichment as the birthplaces of halo stars did.  
This has  already been suggested for the ultra-faint dwarf galaxies \citep{fsg+10}. 
Furthermore, because of the similar CEMP fractions, the origin of CEMP stars in the halo may also lie within early analogs of the surviving dwarf galaxies.

However, we do find that none of our CEMP stars have [C/Fe] $> 1.0$, whereas 32$\%$ of stars in the halo with [Fe/H] $< -3.0$ have [C/Fe] $>$ 1.0 \citep{pfb+14}. 
This discrepancy implies that the distribution and magnitude of carbon-enhancement of CEMP stars in the halo may be different from that in Sculptor.
Thus, while our result does indicate some level of similarity in early chemical enrichment among Sculptor and the Milky Way halo in terms of the CEMP fraction, there may be a level of inhomogeneity in producing the most carbon-enhanced stars.
More observations of Sculptor will further confirm or refute our findings and shed more light on the enrichment history of this galaxy.

Finally, it is interesting to note that the vast majority of CEMP stars in the Milky Way halo with 
[Fe/H] $< -3.0$ are CEMP-no stars, which are stars that display no enhancement in neutron-capture elements.
If our population of CEMP-s candidates have [Fe/H] $< -2.90$, this sample might suggest a discrepancy between the halo and Sculptor in the occurrence rate of CEMP-s stars at low metallicities. However, all of the CEMP-s candidates have only lower limits on their metallicities since strong carbon features blue-ward of the Ca II K line preclude an accurate metallicity measurement. Additional observations with higher resolution spectrographs are needed to verify whether any of our CEMP-s candidates may be EMP stars, although it is unlikely.

Given that most of our CEMP-no candidates have [Fe/H] $< -2.8$, the previous scarcity of CEMP stars in 
Sculptor can likely be explained by the overall rarity of EMP stars in Sculptor and the correspondingly small 
stellar samples at the lowest metallicities with available [Fe/H] and [C/Fe]
measurements. The previously known sample with simultaneous [Fe/H] and [C/Fe] abundances includes 
198 medium-resolution measurements from \citet{kgz+15}, 94 medium-resolution measurements from
\citet{lbp+16}, and 28 stars with high-resolution measurements 
\citep{svt+03, gsw+05, fks+10, tjh+10, kc+12, sht+13, sts+15,
jnm+15,sjf+15}. Thirteen of these stars have [Fe/H] $< -2.8$, one of which is potentially 
a CEMP-no star (Scl11\_1\_4296 in \citet{sjf+15}). This difference (i.e., a low CEMP fraction),
is likely the result of samples that did not target EMP stars systematically as was done in our
IMACS survey or potentially unaccounted for sample biases. Regardless, our sample demonstrates the existence of a substantial population of CEMP stars with [Fe/H]$< -2.8$ in Sculptor.

In summary,  we identified EMP stars in an IMACS survey (Hansen et al. in prep) and based on M2FS follow-up observations,
increased the number of known metal-poor stars in Sculptor with available [Fe/H] and [C/Fe] measurements.
As a result, we provide the first meaningful sample of EMP stars from which 
to determine CEMP fractions to learn about early chemical enrichment and evolution.
Given the similarity to the halo, perhaps all dwarf galaxies share certain properties of early chemical evolution.
Follow-up spectroscopy of additional EMP candidates from 
our IMACS survey will likely lead to even more EMP and CEMP star discoveries in other dwarf galaxies in the future.

\acknowledgements
We thank Joshua Adams for observing and reducing most of the MagE spectra
and Gary da Costa for providing access to the photometric catalog for Sculptor on which these observations were based. 
We would also like to thank Vinicius Placco for computing carbon corrections for this paper. A.C. and A.F. are supported by NSF CAREER grant AST- 1255160. A.C acknowledges support by 
the Whiteman Fellowship at MIT. J.D.S acknowledges support from
AST-1108811. A.F. acknowledges partial support 
from the Silverman (1968) Family Career Development Professorship and PHY 08-22648; Physics Frontier Center/Joint Institute for 
Nuclear Astrophysics (JINA) and PHY 14-30152; and Physics Frontier Center/JINA 
Center for the Evolution of the Elements (JINA-CEE), awarded by the US National 
Science Foundation. This work made use of NASA's Astrophysics Data System Bibliographic Services. M.G.W. is supported by National Science Foundation grants AST-1313045, AST-1412999.
This work has made extensive use of the astropy package \citep{astropy}.

Funding for SDSS-III has been provided by the Alfred P. Sloan Foundation, the Participating Institutions, the National Science Foundation, and the U.S. Department of Energy Office of Science. The SDSS-III web site is http://www.sdss3.org/.

SDSS-III is managed by the Astrophysical Research Consortium for the Participating Institutions of the SDSS-III Collaboration including the University of Arizona, the Brazilian Participation Group, Brookhaven National Laboratory, Carnegie Mellon University, University of Florida, the French Participation Group, the German Participation Group, Harvard University, the Instituto de Astrofisica de Canarias, the Michigan State/Notre Dame/JINA Participation Group, Johns Hopkins University, Lawrence Berkeley National Laboratory, Max Planck Institute for Astrophysics, Max Planck Institute for Extraterrestrial Physics, New Mexico State University, New York University, Ohio State University, Pennsylvania State University, University of Portsmouth, Princeton University, the Spanish Participation Group, University of Tokyo, University of Utah, Vanderbilt University, University of Virginia, University of Washington, and Yale University.

\indent Facilities: Magellan-Clay (M2FS), Magellan-Baade (MagE, IMACS).

\bibliography{scl}

\clearpage
\LongTables
\begin{deluxetable*}{lllllccrccr}
\tablecolumns{11}
\tablewidth{\textwidth}
\tablecaption{M2FS Measurements}
\tablehead{   
  \colhead{Names} &
  \colhead{$\alpha$} &
  \colhead{$\delta$} &
  \colhead{Log (g)} &
  \colhead{Teff} &
  \colhead{[Fe/H]} &
  \colhead{[Fe/H]$_{\text{err}}$} & 
   \colhead{[C/Fe]} &
   \colhead{[C/Fe]$_{\text{err}}$} & 
   \colhead{[C/Fe]$_{\text{correction}}$} &
   \colhead{[C/Fe]$_{\text{final}}$} \\
   	\colhead{} &
  \colhead{(J2000)} &
  \colhead{(J2000)} &
  \colhead{(dex)} &
  \colhead{(K)} &
  \colhead{(dex)} &
  \colhead{(dex)} & 
  \colhead{(dex)} &
  \colhead{(dex)} & 
  \colhead{(dex)} &
  \colhead{(dex)} 
}
\startdata
\multicolumn{5}{}{} & RGB \\
\multicolumn{5}{}{} & members \\
\hline
\\
7\_4\_3266 & 00:58:38.77 & $-$33:35:02.28 & 0.98 & 4461 & $-$2.40 & 0.15 & $-$1.07 & 0.33 & 0.79 & $-0.28$\\
11\_2\_956 & 00:58:39.65 & $-$33:55:34.76 & 1.02 & 4477 & $-$2.16 & 0.18 & $-$0.86 & 0.24 & 0.75 & $-0.11$ \\
7\_4\_3182 & 00:58:49.80 & $-$33:37:19.10 & 1.61 & 4719 & $-$3.05 & 0.26 & 0.07 & 0.37 & 0.3 & $0.37$ \\
11\_1\_6533 & 00:58:57.88 & $-$33:41:50.34 & 1.43 & 4635 & $-$2.98 & 0.22 & 0.09 & 0.44 & 0.48 & $0.57$ \\
11\_1\_6443 & 00:59:00.28 & $-$33:43:14.64 & 1.03 & 4484 & $-$2.45 & 0.17 & $-$0.72 & 0.34 & 0.77 & $0.05$ \\
11\_1\_6267 & 00:59:04.05 & $-$33:40:31.48 & 1.08 & 4503 & $-$2.57 & 0.18 & $-$0.74 & 0.37 & 0.75 & $0.01$ \\
11\_1\_6192 & 00:59:06.14 & $-$33:44:11.39 & 1.37 & 4552 & $-$2.02 & 0.22 & $-$0.89 & 0.26 & 0.57 & $-0.32$ \\
7\_4\_2750 & 00:59:17.20 & $-$33:38:06.68 & 1.36 & 4606 & $-$3.05 & 0.2 & $-$0.34 & 0.42 & 0.55 & $0.21$ \\
11\_2\_661 & 00:59:25.63 & $-$33:58:21.42 & 1.18 & 4524 & $-$3.10 & 0.16 & $-$0.05 & 0.31 & 0.68 & $0.63$ \\
11\_1\_5047 & 00:59:26.68 & $-$33:40:22.43 & 1.49 & 4662 & $-$3.23 & 0.2 & $-$0.01 & 0.35 & 0.36 & $0.35$ \\
7\_4\_2408 & 00:59:30.43 & $-$33:36:05.23 & 1.07 & 4500 & $-$2.68 & 0.16 & $-$0.72 & 0.35 & 0.75 & $0.03$ \\
11\_1\_4824 & 00:59:30.49 & $-$33:39:04.16 & 1.09 & 4508 & $-$2.66 & 0.24 & $-$0.97 & 0.47 & 0.75 & $-0.22$ \\
11\_1\_4673 & 00:59:33.63 & $-$33:49:10.10 & 1.23 & 4546 & $-$3.11 & 0.17 & $-$0.11 & 0.35 & 0.64 & $0.53$ \\
11\_1\_4422 & 00:59:36.61 & $-$33:40:38.51 & 1.76 & 4783 & $-$3.04 & 0.25 & 0.74 & 0.34 & 0.16 & $0.90$ \\
11\_1\_4277 & 00:59:38.42 & $-$33:40:11.57 & 1.81 & 4805 & $-$2.94 & 0.25 & $<0.00$ & ... & 0.12 & $<0.12$ \\
11\_1\_4296 & 00:59:38.75 & $-$33:46:14.58 & 1.55 & 4697 & $-$3.33 & 0.22 & 0.25 & 0.32 & 0.34 & $0.59$ \\
11\_1\_4122 & 00:59:41.24 & $-$33:48:03.56 & 1.2 & 4467 & $-$2.01 & 0.2 & $-$0.88 & 0.24 & 0.67 & $-0.21$ \\
11\_1\_3738 & 00:59:45.30 & $-$33:43:53.83 & 1.79 & 4756 & $-$1.92 & 0.35 & $-$1.01 & 0.37 & 0.26 & $-0.75$ \\
11\_1\_3743 & 00:59:45.37 & $-$33:45:34.19 & 1.66 & 4740 & $-$2.97 & 0.23 & 0.53 & 0.36 & 0.26 & $0.79$ \\
11\_1\_3646 & 00:59:46.67 & $-$33:47:19.71 & 1.72 & 4764 & $-$3.05 & 0.24 & 0.55 & 0.38 & 0.2 & $0.75$ \\
11\_1\_3513 & 00:59:48.19 & $-$33:46:50.01 & 1.59 & 4724 & $-$2.62 & 0.27 & 0.18 & 0.39 & 0.37 & $0.55$ \\
11\_2\_425 & 00:59:50.64 & $-$33:58:07.10 & 1.6 & 4715 & $-$3.15 & 0.22 & 0.39 & 0.37 & 0.33 & $0.72$ \\
7\_3\_243 & 00:59:50.78 & $-$33:31:47.06 & 1.25 & 4491 & $-$1.48 & 0.27 & $-$1.32 & 0.32 & 0.66 & $-0.66$ \\
11\_1\_3246 & 00:59:51.19 & $-$33:44:51.82 & 1.36 & 4546 & $-$1.83 & 0.58 & $-$1.05 & 0.53 & 0.58 & $-0.47$ \\
10\_8\_4250 & 00:59:51.51 & $-$33:44:02.67 & 1.29 & 4573 & $-$2.73 & 0.22 & $-$0.75 & 0.40 & 0.63 & $-0.12$ \\
7\_4\_1514 & 00:59:54.47 & $-$33:37:53.50 & 1.23 & 4479 & $-$1.45 & 0.26 & $-$1.14 & 0.27 & 0.64 & $-0.50$ \\
10\_8\_4020 & 00:59:55.22 & $-$33:42:11.34 & 1.4 & 4624 & $-$3.05 & 0.21 & $-$0.07 & 0.36 & 0.51 & $0.44$ \\
11\_1\_2583 & 00:59:57.59 & $-$33:38:32.54 & 1.35 & 4539 & $-$1.78 & 0.78 & $-$0.78 & 0.61 & 0.56 & $-0.22$ \\
6\_5\_1598 & 00:59:59.09 & $-$33:36:44.90 & 1.09 & 4492 & $-$2.92 & 0.16 & 0.18 & 0.34 & 0.67 & $0.85$ \\
10\_8\_3751 & 00:59:59.33 & $-$33:44:24.34 & 1.6 & 4711 & $-$3.05 & 0.24 & 0.18 & 0.39 & 0.3 & $0.48$ \\
10\_8\_3709 & 00:59:59.95 & $-$33:47:02.03 & 1.67 & 4742 & $-$2.85 & 0.24 & 0.26 & 0.40 & 0.25 & $0.51$ \\
10\_8\_3698 & 01:00:00.04 & $-$33:45:28.81 & 1.18 & 4546 & $-$2.59 & 0.21 & $-$0.47 & 0.38 & 0.69 & $0.22$ \\
10\_7\_923 & 01:00:01.12 & $-$33:59:21.38 & 1.4 & 4641 & $-$3.77 & 0.20 & $<0.34$ & 0.36 & 0.47 & $<0.81$ \\
10\_8\_3625 & 01:00:01.44 & $-$33:51:16.74 & 1.0 & 4469 & $-$2.11 & 0.17 & $-$0.84 & 0.31 & 0.75 & $-0.09$ \\
10\_8\_3520 & 01:00:03.27 & $-$33:47:44.44 & 1.33 & 4591 & $-$2.85 & 0.21 & $-$0.38 & 0.38 & 0.59 & $0.21$ \\
10\_8\_3315 & 01:00:05.93 & $-$33:45:56.39 & 0.99 & 4465 & $-$2.54 & 0.18 & $-$0.59 & 0.35 & 0.76 & $0.17$ \\
10\_8\_3167 & 01:00:07.86 & $-$33:47:07.62 & 1.51 & 4672 & $-$3.05 & 0.22 & 0.11 & 0.42 & 0.4 & $0.51$ \\
10\_8\_2933 & 01:00:11.19 & $-$33:40:38.65 & 1.78 & 4790 & $-$2.96 & 0.23 & $<0.25$ & ... & 0.14 & $<0.39$ \\
10\_8\_2927 & 01:00:11.30 & $-$33:39:35.67 & 1.18 & 4527 & $-$2.94 & 0.17 & 0.03 & 0.39 & 0.65 & $0.68$ \\
10\_8\_2908 & 01:00:11.72 & $-$33:44:50.34 & 0.99 & 4451 & $-$2.78 & 0.15 & $-$0.53 & 0.32 & 0.75 & $0.22$ \\
10\_8\_2824 & 01:00:12.77 & $-$33:38:53.56 & 1.45 & 4646 & $-$3.14 & 0.22 & 0.39 & 0.31 & 0.45 & $0.84$ \\
10\_8\_2818 & 01:00:12.95 & $-$33:42:03.91 & 1.2 & 4532 & $-$2.83 & 0.19 & $-$0.18 & 0.34 & 0.66 & $0.48$ \\
10\_8\_2730 & 01:00:14.49 & $-$33:47:50.49 & 1.35 & 4601 & $-$2.86 & 0.22 & $-$0.37 & 0.40 & 0.57 & $0.20$ \\
10\_8\_2669 & 01:00:15.26 & $-$33:45:49.87 & 1.83 & 4814 & $-$2.94 & 0.23 & $-$0.03 & 0.38 & 0.1 & $0.07$ \\
10\_8\_2647 & 01:00:15.67 & $-$33:45:59.96 & 1.49 & 4680 & $-$2.39 & 0.29 & $-$0.11 & 0.34 & 0.52 & $0.41$ \\
10\_8\_2635 & 01:00:15.87 & $-$33:45:01.90 & 1.39 & 4616 & $-$3.05 & 0.2 & $-$0.28 & 0.36 & 0.52 & $0.24$ \\
10\_8\_2558 & 01:00:17.03 & $-$33:42:47.26 & 1.88 & 4837 & $-$2.91 & 0.28 & 0.22 & 0.47 & 0.09 & $0.31$ \\
6\_5\_1035 & 01:00:19.33 & $-$33:37:11.74 & 1.27 & 4564 & $-$3.03 & 0.2 & $-$0.30 & 0.35 & 0.62 & $0.32$ \\
6\_5\_948 & 01:00:22.37 & $-$33:38:07.79 & 1.39 & 4633 & $-$2.50 & 0.27 & $-$0.18 & 0.35 & 0.57 & $0.39$ \\
10\_8\_2211 & 01:00:22.74 & $-$33:51:22.84 & 1.18 & 4456 & $-$1.59 & 0.25 & $-$1.09 & 0.27 & 0.65 & $-0.44$ \\
10\_8\_2148 & 01:00:23.49 & $-$33:41:46.18 & 1.76 & 4785 & $-$2.83 & 0.27 & 0.55 & 0.35 & 0.18 & $0.73$ \\
10\_8\_2126 & 01:00:24.07 & $-$33:45:54.41 & 1.4 & 4620 & $-$2.74 & 0.24 & 0.38 & 0.36 & 0.46 & $0.84$ \\
10\_8\_2028 & 01:00:25.95 & $-$33:48:40.71 & 1.53 & 4697 & $-$2.39 & 0.51 & $-$0.22 & 0.51 & 0.47 & $0.25$ \\
10\_8\_1887 & 01:00:28.43 & $-$33:47:41.51 & 1.19 & 4530 & $-$2.72 & 0.21 & $-$0.26 & 0.36 & 0.68 & $0.42$ \\
10\_8\_1877 & 01:00:28.63 & $-$33:46:02.64 & 1.49 & 4607 & $-$1.87 & 0.26 & $-$0.70 & 0.30 & 0.46 & $-0.24$ \\
10\_8\_1731 & 01:00:31.00 & $-$33:47:12.23 & 1.96 & 4869 & $-$2.91 & 0.23 & $<0.25$ & ... & 0.03 & $<0.28$ \\
6\_5\_736 & 01:00:31.87 & $-$33:38:00.22 & 1.23 & 4547 & $-$3.03 & 0.18 & $-$0.07 & 0.38 & 0.63 & $0.56$ \\
10\_8\_1640 & 01:00:32.68 & $-$33:41:05.05 & 1.8 & 4758 & $-$1.59 & 0.26 & $-$0.89 & 0.27 & 0.3 & $-0.59$ \\
10\_8\_1566 & 01:00:33.94 & $-$33:40:08.24 & 1.04 & 4486 & $-$2.42 & 0.2 & $-$0.74 & 0.31 & 0.77 & $0.03$ \\
6\_5\_678 & 01:00:34.10 & $-$33:35:08.73 & 1.38 & 4615 & $-$2.74 & 0.24 & 0.09 & 0.43 & 0.53 & $0.62$ \\
10\_7\_570 & 01:00:36.41 & $-$33:52:19.54 & 1.81 & 4805 & $-$2.94 & 0.24 & 0.37 & 0.34 & 0.13 & $0.50$ \\
10\_8\_1463 & 01:00:36.46 & $-$33:50:26.67 & 1.96 & 4871 & $-$2.91 & 0.27 & 0.22 & 0.32 & 0.04 & $0.26$ \\
10\_8\_1325 & 01:00:39.72 & $-$33:39:12.42 & 2.03 & 4870 & $-$1.96 & 0.48 & $-$0.53 & 0.41 & 0.04 & $-0.49$ \\
10\_8\_1308 & 01:00:40.35 & $-$33:44:14.23 & 1.36 & 4603 & $-$2.97 & 0.23 & $-$0.01 & 0.40 & 0.54 & $0.53$ \\
10\_8\_1124 & 01:00:46.21 & $-$33:42:34.03 & 1.21 & 4539 & $-$2.72 & 0.2 & $-$0.49 & 0.38 & 0.68 & $0.19$ \\
10\_8\_1072 & 01:00:47.83 & $-$33:41:03.17 & 1.3 & 4581 & $-$3.63 & 0.21 & $<0.25$ & ... & 0.56 & $<0.81$ \\
10\_8\_1062 & 01:00:48.14 & $-$33:42:13.32 & 1.93 & 4859 & $-$2.91 & 0.23 & 0.37 & 0.30 & 0.06 & $0.43$ \\
10\_7\_442 & 01:00:50.35 & $-$33:52:15.67 & 1.62 & 4723 & $-$3.15 & 0.22 & 0.39 & 0.38 & 0.29 & $0.68$ \\
6\_5\_420 & 01:00:51.64 & $-$33:36:56.74 & 2.5 & 4565 & $-$0.61 & 0.46 & $-$0.40 & 0.40 & 0.03 & $-0.37$  \\
10\_8\_798 & 01:00:56.41 & $-$33:49:47.18 & 1.37 & 4609 & $-$2.74 & 0.23 & $-$0.34 & 0.37 & 0.55 & $0.21$ \\
10\_8\_758 & 01:00:57.56 & $-$33:39:39.74 & 1.64 & 4746 & $-$2.50 & 0.28 & 0.00 & 0.38 & 0.32 & $0.32$ \\
10\_8\_577 & 01:01:06.92 & $-$33:46:13.15 & 1.74 & 4773 & $-$3.23 & 0.27 & 0.61 & 0.39 & 0.17 & $0.78$ \\
6\_5\_239 & 01:01:10.27 & $-$33:38:37.81 & 1.09 & 4505 & $-$2.44 & 0.23 & $-$0.43 & 0.32 & 0.72 & $0.29$ \\
10\_8\_462 & 01:01:13.19 & $-$33:43:20.56 & 1.53 & 4681 & $-$3.06 & 0.22 & 0.28 & 0.36 & 0.38 & $0.66$ \\
10\_8\_320 & 01:01:22.24 & $-$33:46:21.81 & 1.1 & 4493 & $-$3.00 & 0.15 & $-$0.39 & 0.40 & 0.72 & $0.33$ \\
10\_8\_265 & 01:01:27.22 & $-$33:45:15.31 & 1.51 & 4671 & $-$3.05 & 0.22 & 0.37 & 0.34 & 0.41 & $0.78$ \\
10\_8\_61 & 01:01:47.52 & $-$33:47:27.64 & 1.6 & 4713 & $-$3.15 & 0.25 & $<0.75$ & ... & 0.3 & $<1.05$ \\
\hline
\\
\multicolumn{5}{}{} & CEMP-s \\
\multicolumn{5}{}{} & candidates \\
\hline
\\
11\_1\_6440$^\dagger$ & 00:59:00.13 & $-$33:38:50.96 & 1.3 & 4579 & $>-$3.04 & ... & ... & ... & ... & ... \\
11\_1\_5437$^\dagger$ & 00:59:19.87 & $-$33:38:56.77 & 1.12 & 4517 & $>-$3.41 & ... & ... & ... & ... & ... \\
11\_1\_4121$^\dagger$ & 00:59:41.05 & $-$33:45:25.28 & 1.25 & 4554 & $>-$3.12 & ... & ... & ... & ... & ... \\
11\_1\_3334$^\dagger$ & 00:59:49.62 & $-$33:40:41.78 & 1.62 & 4721 & $>-$3.24 & ... & ... & ... & ... & ... \\
10\_8\_3963$^\dagger$ & 00:59:56.17 & $-$33:43:04.89 & 1.09 & 4488 & $>-$3.09 & ... & ... & ... & ... & ... \\
10\_8\_3926$^\dagger$ & 00:59:56.73 & $-$33:39:37.54 & 1.36 & 4626 & $>-$3.76 & ... & ... & ... & ... & ... \\
10\_8\_3804$^\dagger$ & 00:59:58.91 & $-$33:50:53.61 & 1.63 & 4727 & $>-$3.15 & ... & ... & ... & ... & ... \\
10\_8\_2134$^\dagger$ & 01:00:23.71 & $-$33:40:20.40 & 1.41 & 4628 & $>-$2.98 & ... & ... & ... & ... & ... \\
10\_7\_486$^\dagger$ & 01:00:45.41 & $-$33:52:14.68 & 1.14 & 4509 & $>-$3.02 & ... & ... & ... & ... & ... \\
6\_5\_505$^\dagger$ & 01:00:45.76 & $-$33:38:34.83 & 1.57 & 4706 & $>-$3.33 & ... & ... & ... & ... & ... \\
10\_8\_437$^\dagger$ & 01:01:15.05 & $-$33:50:02.63 & 1.24 & 4553 & $>-$3.20 & ... & ... & ... & ... & ... \\
\hline
\\
\multicolumn{5}{}{} & Blueward \\
\multicolumn{5}{}{} & of RGB \\
\hline
\\
10\_8\_4247 & 00:59:51.56 & $-$33:45:07.76 & 3.32 & 5419 & $-$2.84 & 0.39 & $<0.40$ & ... & 0.0 & $<0.40$ \\
10\_8\_4014 & 00:59:55.48 & $-$33:45:51.48 & 2.88 & 5244 & $-$2.99 & 0.42 & 0.64 & 0.51 & 0.01 & $0.65$ \\
10\_8\_3723 & 00:59:59.92 & $-$33:51:11.79 & 2.96 & 5286 & $-$2.38 & 0.82 & $<0.40$ & ... & 0.01 & $<0.41$ \\
10\_8\_3558 & 01:00:02.65 & $-$33:49:18.73 & 3.02 & 5309 & $-$2.48 & 0.31 & $<0.4$ & ... & 0.01 & $<0.41$ \\
10\_8\_3188 & 01:00:07.66 & $-$33:49:46.99 & 3.32 & 5394 & $-$1.98 & 0.45 & $<0.00$ & ... & 0.0 & $<0.00$ \\
10\_8\_3111 & 01:00:08.86 & $-$33:49:49.67 & 2.65 & 5066 & $-$1.15 & 0.37 & $-$0.86 & 0.30 & 0.02 & $-0.84$ \\
10\_8\_3045 & 01:00:09.72 & $-$33:47:00.79 & 2.49 & 5100 & $-$2.55 & 0.28 & 0.16 & 0.39 & 0.01 & $0.17$ \\
10\_8\_1615 & 01:00:33.05 & $-$33:43:02.26 & 2.61 & 5151 & $-$2.55 & 0.34 & $<0.25$ & ... & 0.01 & $<0.26$ \\
10\_8\_1366 & 01:00:38.71 & $-$33:43:16.58 & 2.81 & 5227 & $-$2.10 & 0.38 & $-$0.22 & 0.41 & 0.01 & $-0.21$ \\
10\_8\_440 & 01:01:14.29 & $-$33:39:27.82 & 3.59 & 5493 & $-$1.65 & 0.2 & $-$0.45 & 0.31 & 0.0 & $-0.45$ \\
10\_8\_436 & 01:01:14.95 & $-$33:47:21.34 & 3.35 & 5404 & $-$1.42 & 0.83 & $-$0.47 & 0.57 & 0.0 & $-0.47$ \\
6\_5\_163 & 01:01:19.89 & $-$33:35:57.44 & 2.41 & 5055 & $-$2.95 & 0.45 & 0.84 & 0.48 & 0.01 & $0.85$

\enddata
\tablecomments{Stars in the upper section lie on the RGB of Sculptor, and stars in the lower section lie blueward of the RGB (see Figure~\ref{fig:CMD}).\\
$^\dagger$ These stars are classified as likely CH-strong, Ba-strong, or CEMP-s stars due to the presence of saturated carbon features (see Sections~\ref{sec:crich} and~\ref{sec:bafe}).}
\label{tab:M2FS}
\end{deluxetable*}


\end{document}